\documentclass[prd,superscriptaddress,amsfonts,amssymb,amsmath,twocolumn,floatfix]{revtex4-2}
\let\revappendix\appendix
\usepackage{bm}
\usepackage{amsfonts}
\usepackage{latexsym}
\usepackage{libertine}
\usepackage{graphicx}
\usepackage{amsmath}
\usepackage{braket}
\usepackage{palatino}
\usepackage{bigints}
\usepackage{mathpazo}
\usepackage{textcomp}
\usepackage{float}
\usepackage{booktabs}
\usepackage{dcolumn}
\usepackage{booktabs}
\usepackage{multirow}
\usepackage{hyperref}
\hypersetup{colorlinks,citecolor=blue}
\usepackage{amsmath}
\usepackage{xcolor}
\usepackage{orcidlink}
\usepackage{commath}
\usepackage{subcaption}

\def\jnl@style{\it}
\def\aaref@jnl#1{{\jnl@style#1}}

\def\aaref@jnl#1{{\jnl@style#1}}

\def\aj{\aaref@jnl{AJ}}                   
\def\apj{\aaref@jnl{ApJ}}                 
\def\apjl{\aaref@jnl{ApJ}}                
\def\apjs{\aaref@jnl{ApJS}}               
\def\apss{\aaref@jnl{Ap\&SS}}             
\def\aap{\aaref@jnl{A\&A}}                
\def\aapr{\aaref@jnl{A\&A~Rev.}}          
\def\aaps{\aaref@jnl{A\&AS}}              
\def\mnras{\aaref@jnl{Mon.~Not.~Roy.~Astron.~Soc.}}             
\def\prd{\aaref@jnl{Phys.~Rev.~D}}        
\def\prc{\aaref@jnl{Phys.~Rev.~C}}  
\def\prl{\aaref@jnl{Phys.~Rev.~Lett.}}    
\def\qjras{\aaref@jnl{QJRAS}}             
\def\skytel{\aaref@jnl{S\&T}}             
\def\ssr{\aaref@jnl{Space~Sci.~Rev.}}     
\def\zap{\aaref@jnl{ZAp}}                 
\def\nat{\aaref@jnl{Nature}}              
\def\aplett{\aaref@jnl{Astrophys.~Lett.}} 
\def\apspr{\aaref@jnl{Astrophys.~Space~Phys.~Res.}} 
\def\physrep{\aaref@jnl{Phys.~Rep.}}      
\def\physscr{\aaref@jnl{Phys.~Scr}}       
\def\commat{\aaref@jnl{Comm.~Math.~Phys.}}              
\def\science{\aaref@jnl{Science}}               
\def\cqg{\aaref@jnl{Classical Quant.~Grav.}}            
\def\jpcs{\aaref@jnl{JPCS}}                                     
\def\ijmpd{\aaref@jnl{Int.~J.~Mod.~Phys.~D}}                    
\def\grg{\aaref@jnl{Gen.~Relat.~Gravit.}}               
\def\rpp{\aaref@jnl{Rep.~Prog.~Phys.}}          
\def\npa{\aaref@jnl{Nucl.~Phys.~A}}        
\def\lrr{\aaref@jnl{Living Rev.~Rel.}}                   
\def\jcap{\aaref@jnl{J.~Cosmology Astropart.~Phys.}}    
\def\rmp{\aaref@jnl{Rev.~Mod.~Phys.}}   
\def\epjc{\aaref@jnl{Eur.~Phys.~J.~C}}

\allowdisplaybreaks[1]

\begin{document}

\color{black}       

\title{Four $\mathbb{S}\mathbb{T}\mathbb{U}$ Black Holes Shadows}
\author{Yassine Sekhmani\orcidlink{0000-0001-7448-4579}}
\email[Email: ]{sekhmaniyassine@gmail.com (corresponding author)}
\affiliation{Ratbay Myrzakulov Eurasian International Centre for Theoretical Physics, Astana 010009, Kazakhstan.}
\affiliation{L. N. Gumilyov Eurasian National University, Astana 010008,
Kazakhstan.}

\author{Dhruba Jyoti Gogoi\orcidlink{0000-0002-4776-8506}
}
\email[Email: ]{moloydhruba@yahoo.in}
\affiliation{Department of Physics, Moran College, Moranhat, Charaideo 785670, Assam, India.}
\affiliation{Theoretical Physics Division, Centre for Atmospheric Studies, Dibrugarh University, Dibrugarh
786004, Assam, India.}

\author{Ratbay Myrzakulov\orcidlink{0000-0002-5274-0815}}
\email[Email:]{rmyrzakulov@gmail.com }
\affiliation{Ratbay Myrzakulov Eurasian International Centre for Theoretical
Physics, Astana 010009, Kazakhstan.}
\affiliation{L. N. Gumilyov Eurasian National University, Astana 010008,
Kazakhstan.}

\author{Giuseppe Gaetano Luciano\orcidlink{0000-0002-5129-848X}
}
\email[Email: ]{giuseppegaetano.luciano@udl.cat}
\affiliation{Department of Chemistry, Physics and Environmental and Soil Sciences, Escola Politecninca Superior, Universidad de Lleida, Av. Jaume
II, 69, 25001 Lleida, Spain.}

\author{Javlon Rayimbaev\orcidlink{0000-0001-9293-1838}}
\email[Email: ]{javlon@astrin.uz}
\affiliation{
New Uzbekistan University, Movarounnahr St. 1, Tashkent 100007, Uzbekistan}
\affiliation{
University of Tashkent for Applied Sciences, Gavhar Str. 1, Tashkent 100149, Uzbekistan }
\affiliation{National University of Uzbekistan, Tashkent 100174, Uzbekistan}

\begin{abstract}
In this work, we examine the optical behaviors and thermodynamic phase structures using shadow analysis for four $\mathbb{S}\mathbb{T}\mathbb{U}$ black holes. The study is conducted for four cases of charge configurations on the parameter space $\mathcal{M}\left(q_1,q_2,q_3,q_4\right)$. As a matter of fact, both the electric charge as a parameter and the parameter space $\mathcal{M}\left(q_1,q_2,q_3,q_4\right)$ affect the geometry of the black hole shadow, particularly the size of the shadow. We also introduce a constraint on the charge of the black hole from the observational results of the M87$^\star$ {\color{black}and Sgr A$^\star$} shadow. Furthermore, we show that the electric charge and the parameter space $\mathcal{M}\left(q_1,q_2,q_3,q_4\right)$ have a non-trivial impact on the variation of the energy emission rate. Interestingly enough, we find novel scenarios in which the evaporation is slower, which causes the lifetime of the black holes to be considerably elongated. On the other side, the phase structure of four $\mathbb{S}\mathbb{T}\mathbb{U}$ black holes is explored for two cases of electric charge configuration. The findings show a perfect correlation between the shadow and event horizon radii. This correlation is, in fact, helpful in discovering the phase transition in terms of the shadow radius. In addition, the microstructure is being analyzed in terms of shadow analysis, providing similar behavior to the ordinary situation of the Ruppeiner formalism.
\end{abstract}


\maketitle

\section{Introduction}
The optical appearance of black holes has recently received special attention thanks to the first-ever event-horizon-scale images of M87$^\star$ \cite{1,2,3,4,5,6} and Sgr A$^\star$ \cite{7}. These observations have attracted a great deal of interest in the context of general relativity (GR) and modified gravity (MOG), offering the possibility of showing and examining specific mimetic images related to the observed images. This possibility allows us to discuss, from a geometric or topological point of view, a comparison between the specific theory of the black hole image and the observed images of M87$^\star$ or Sgr A$^\star$. From this perspective, the black hole image becomes an interesting optical aspect of the black hole. In particular, the attempt to discover the relationship between optical properties, namely shadow behaviors, is carried out with thermodynamics \cite{8,9} and quasi-normal modes \cite{10,11} with the aim of unifying the physical interpretation. In addition, from the point of view of shape data, the shadow image is either a set of geometries distorted on the image plane \cite{12,13,14,15,16,Zahid:2023csk,Khan:2023qsl}, or a regular set of concentric circles \cite{Rayimbaev:2022hca}, depending on whether or not the space-time is endowed with the rotation parameter. For a better understanding of shadow behaviors, the Hamilton-Jacobi formalism \cite{17} is the compatible analytical structure based on the fact that a massless photon near a black hole generates such orbits around the black hole region, also known as the geodesic null background. In the context of black hole imaging, research is underway to provide a comparative study between the observed images, namely M87$^\star$ and Sgr A$^\star$, and the theoretical image with regard to size and shape. The latest shadow-related activities are generally fruitful in theoretical physics; in particular, Einstein-Maxwell dilaton-axion gravity (EMDA) produces a relevant image of the black hole shadow with a set of appropriate parameter spaces, i.e. the spin parameter and the dilaton parameter \cite{18}. A parallel study has attempted to reveal the impact of the axionic coupling parameter on black hole shadows in modified Chern-Simons gravity \cite{19}. On the other hand, black hole shadow and chaos bound violation have been recently discussed in \cite{199} in f(T) teleparallel gravity.

The pioneering study of the thermodynamic phase transition of black holes has led to an in-depth understanding of the hidden structure of these compact objects, giving rise to multiple types of critical behavior. In particular, the thermodynamic analysis of critical behaviors of charged AdS black holes is conducted in an \emph{extended phase space} by identifying the negative cosmological constant as thermodynamic pressure and the black hole mass with the enthalpy. The ensuing framework is typically referred to as black hole chemistry \cite{20,21}. For recent applications, see also \cite{21a,21b,21c,21d,21e,21f}. 
 On the other hand, in the spacetime geometry, the Hawking page phase transition \cite{22} involves a transition between Anti-de Sitter (AdS) black holes with radiation and AdS thermal. To this end, several works have attempted to interpret the thermodynamic characteristics of the black hole from the point of view of a shadow background. The expected result shows a positive correlation between horizon radius and shadow radius \cite{23,24}. This feature is powerful evidence to support the study of thermodynamic black holes in terms of shadow analysis.

{\color{black}The AdS black hole solutions of gauged extended supergravity \cite{R1} are of great interest \cite{Behrndt:1998ns,Birmingham:1998nr,Caldarelli:1998hg,Klemm:1998in,Behrndt:1998jd,Duff:1999gh,Chamblin:1999tk,Cvetic:1999ne,Cvetic:1999rb,Sabra:1999ux,Caldarelli:1999ar}, mainly owing to the connection between AdS and conformal field theories along the boundary \cite{Maldacena:1997re,Gubser:1998bc,Witten:1998qj,Witten:1998zw}. These gauged extended supergravities might emerge as the massless modes of miscellaneous Kaluza-Klein compactifications of the $D = 11$ supergravity. One of the gauged theories that will be particularly discussed in this paper is the $D = 4$, $N = 8$, $\textsc{SO}(8)$ gauged supergravity \cite{deWit:1981sst,deWit:1982bul} stemming from the $D = 11$ supergravity on $S^7$~\cite{Duff:1986hr}. \textcolor{black}{Specifically, the $N=2, D=4$ gauging under consideration is the $\textsc{U}(1) \in \textsc{SU}(2)_R$ Fayet-Iliopoulos gauging (see~\cite{Bellucci2} for details).}
Within this model, the solutions for black holes are described in \cite{Duff:1999gh}. Without black holes, this AdS compactification is identified as originating from the near-horizon geometry of the non-rotating $M2$ extremal brane \cite{Gibbons:1993sv,Duff:1994fg}. As these gauged supergravity theories may be consistently derived using truncation of the massive modes of full Kaluza-Klein theories, it immediately follows that any solutions of lower-dimensional theories will likewise be solutions of higher-dimensional theories \cite{Duff:1985jd,Pope:1984his}. Basically, having mastered the Kaluza-Klein ansatz for the massless sector, it should be easy to interpret higher-dimensional solutions. In this respect, it is quite adequate to consider truncations of gauged supergravities to include only gauge fields in Cartan subalgebras of full gauge groups, i.e. $\textsc{U}(1)^4$ for $S^7$ compactification. This truncated theory is said to admit $4$-charge $AdS_4$ black hole solutions.

In four dimensions, there exists a consistent truncation of the maximal gauged supergravity $N = 8$ to a gauged supergravity $N = 2$ coupled to three vector multiplets. Accordingly, the corresponding bosonic content of truncated theory $N = 2$ consists of a graviton, four vectors (one of which corresponding to a graviphoton, while the remaining three to the vector multiplets), as well as three complex scalars, whose real and imaginary parts amount to three "axions" and three "dilatons". The involvement of axions is quite essential to providing a coherent truncation. However, the embedding in an alternative truncation in which all three axionic scalar fields are set to zero has been reported in \cite{Duff:1999gh}. These truncations correspond to the Cartan $\textsc{U}(1)^4$ subgroup of the non-abelian $\textsc{SO}(8)$, for which there are AdS black hole solutions with four electric charges. 
Practically speaking, the four vectors are the two Kaluza-Klein gauge fields and the two winding gauge fields, whilst the three complex scalars  $\mathbb{S}$, $\mathbb{T}$ and $\mathbb{U}$ stand for the axion-dilaton, the K\"{a}hler shape and the complex structure of the torus. Broadly conceived, this $\mathbb{S}\mathbb{T}\mathbb{U}$ system holds a prominent role in four-dimensional string/string/string triality \cite{Duff:1995sm,Behrndt:1996hu,Bellucci:2008sv,Gimon,Belhaj:2017yva,Belhaj:2016yyq}. Also well known are the black hole solutions \cite{Duff:1995sm, Cvetic:1995uj} of this theory and their embedding in the ungauged supergravity $N = 8$ \cite{Lu:1995yn,Khuri:1995xk}, resulting from the $T^7$ compactification of M-theory in the form of intersections \cite{Cvetic:1995yq,Tseytlin:1996bh}. For further study along this direction, see also~\cite{26,YasLuc}. 
}

To investigate the charge effect on shadow behaviors, the concept of four $\mathbb{S}\mathbb{T}\mathbb{U}$ black holes is employed. In the essence of the $\mathbb{S}\mathbb{T}\mathbb{U}$ black holes, multiple studies of thermodynamics~\cite{Sekhmani:2023qqe,Sekhmani:2023est}, quasi-normal modes or related fields have been investigated. In particular, the extremal charged black holes in $\mathbb{S}\mathbb{T}\mathbb{U}$ supergravity display superradiant instability caused by the existence of unstable (low-frequency) quasi-bond states related to a massive charged scalar field. For that reason, according to \cite{30}, the $\mathbb{S}\mathbb{T}\mathbb{U}$ black holes for certain charge configurations can excite the (higher-frequency) super radiantly unstable quasinormal modes. Moreover, the unstable modes are strictly associated with the volcano-shaped effective potential in the Schrödinger-type wave equation. On the other hand, the holographic entanglement entropy of the $\mathbb{S}\mathbb{T}\mathbb{U}$ black hole \cite{31} is taken into consideration for certain charge configurations. Consequently, in the case where the gravitational background exhibits Van der Waals-like properties, the entanglement entropy shows a transition at a defined critical temperature. In this way, entanglement entropy effectively involves the information of the extended phase structure. Furthermore, the Gauss-Bonnet (GB) quadratic term is the next leading order of the $\alpha'$-expansion of type IIB superstring theory, with $\alpha'$ being the inverse string tension \cite{32,33}. Recently, the subject of the Gauss-Bonnet term has attracted particular interest because of its rescaling into four-dimensional spacetime. In the light of the rescaled Einstein Gauss-Bonnet gravity, numerous investigations have been conducted into thermodynamic behaviors \cite{34,35,36,37}, shadow behaviors \cite{38,39,40,Zubair:2023cep,Rayimbaev:2022znx,Zahid:2022eeq} or quasinormal modes \cite{Sekhmani:2024xyd,42,43} for several black hole solutions.

Starting from the above premises, the present study aims at investigating the shadow behavior and the energy emission rate as optical properties, and the phase structure of the $\mathbb{S}\mathbb{T}\mathbb{U}$ black hole using shadow analysis. In particular, we want to analyze how the electric charge and the parameter space of four electric charges affect the optical and thermodynamic properties. Although this is a toy model analysis, results are shown to be highly non-trivial and provide valuable insights toward the generalization to the more realistic picture including rotational effects. As a feature of the rotating black hole case, it should be remarked that the shadow is a collection of concentric deformed curves that generate the distortion parameter in the study. This deformation is due to the fact that the horizon of the rotating black hole is not spherical. Since this is the major difference with a static black hole, we expect the results in the rotating case not to depart significantly from the present ones. 
Such an upgrade is however under active consideration and will be presented in more detail elsewhere.

The considered paper is organized as follows: In Section $\ref{S2}$, we first briefly introduce the $\mathbb{S}\mathbb{T}\mathbb{U}$ model. In Section $\ref{S3}$, we study shadow behaviors in the $g^{\mu\nu}(x)$ spacetime and the energy emission rate of the $\mathbb{S}\mathbb{T}\mathbb{U}$ black hole with respect to four cases of charge configurations. In Section $\ref{S4}$, we examine a comparative study from the M87$^\star$ constraints. In Section $\ref{S5}$, we study the phase structure of the $\mathbb{S}\mathbb{T}\mathbb{U}$ black hole using shadow analysis, applying the Ruppeiner formalism to investigate the microstructure in terms of the shadow radius. In the conclusion, we thoroughly summarize the results.

\section{Gauged $\mathbb{S}\mathbb{T}\mathbb{U}$  Supergravity}
\label{S2}
{\color{black} 
We look at $D = 4$ theories, which emerge from a truncation of the $N = 8$ gauged supergravity \cite{deWit:1981sst,deWit:1982bul} \textcolor{black}{by imposing invariance under the maximal torus of the pertinent gauge group. The latter is the non-maximal compact subgroup $SO(8)$ of  $\textsc{E}_{7(7)}$, corresponding to the isometry group of $S^7$.
Toward this end, we basically follow the analysis of~\cite{Larios:2023lxg}.}
Here, the bosonic field consists of the metric, four vectors, and six (pseudo) scalars corresponding to the scalar manifold 
\begin{equation}\label{sm}
    \left(\frac{\textsc{SL}(2,\mathbb{R})}{\textsc{SO(2)}}\right)^3\subset\frac{\textsc{E}_{7(7)}}{\textsc{SU(8)}},
\end{equation}
parameterised by $u_i = \chi_i-i e^{-\phi_i}$, with $i = 1, 2, 3$. The bosonic sector of the Lagrangians of these $\mathbb{S}\mathbb{T}\mathbb{U}$ supergravities could in turn be written as follows
\begin{equation}\label{lg}
    \mathcal{L}=\left(R-V\right)\textrm{vol}_4+\mathcal{L}_{\textsc{NLSM}}+\mathcal{L}_{\textsc{vec}}.
\end{equation}
The scalar kinetic terms read
\begin{equation}
    \mathcal{L}_{\textsc{NLSM}}=\frac{1}{2}\sum_i\left(\mathrm{d}\phi_i\wedge\star\mathrm{d}\phi_i+e^{2\phi_i}\mathrm{d}\chi_i\wedge\star\mathrm{d}\chi_i\right),
\end{equation}
and the vector kinetic terms are set by
\begin{equation}
     \mathcal{L}_{\textsc{vec}}=\frac{1}{2}\mathcal{I}_{ab}F^a\wedge\star F^b+\frac{1}{2}\mathcal{R}_{ab}F^a\wedge F^b,
\end{equation}
with $a = 1, 2, 3, 4$. Here $R$ is the scalar curvature. The index $i$ designates the $\phi_i$ dilatons and $\chi_i$ axions according to the range $1 \leq i \leq 3$, while $F^a$ are the field strengths for the four abelian gauge fields $A^a$. These non-minimal couplings can be extracted from the symmetric coset representative of the maximal theory $N=8$ via a block decomposition. The output can be represented as the period matrix $\mathcal{N}_{ab}=\mathcal{R}_{ab}+ i\mathcal{I}_{ab}$ in the form~\cite{Larios:2023lxg}
\begin{widetext}
    \begin{equation}\mathcal{N}_{ab}=\frac{i}{W}
    \begin{pmatrix}
-\Tilde{Y}_1^2\Tilde{Y}_2^2\Tilde{Y}_3^2  & q_1Y_1^2 & q_3\Tilde{Y}_3^2 & q_2Y_2^2\\
q_1Y_1^2 & -\Tilde{Y}_1^2Y_2^2Y_3^2  &  q_2\Tilde{Y}_2^2 & q_3Y_3^2\\
q_3\Tilde{Y}_3^2 & q_2\Tilde{Y}_2^2 & - Y_1^2\Tilde{Y}_2^2Y_3^2 & q_1\Tilde{Y}_1^2\\
q_2Y_2^2 & q_3Y_3^2 & q_1\Tilde{Y}_1^2 & -Y_1^2Y_2^2\Tilde{Y}_3^2
\end{pmatrix}.
\end{equation}
\end{widetext}
Here, we have defined \cite{Azizi:2016noi}
\begin{align}
   \Tilde{Y}_i^2&= e^{-\phi_i}+e^{\phi_i}\chi_i^2,\nonumber\\
   Y_i^2&=e^{\phi_i},\nonumber\\
   b_i&=\chi_ie^{\phi_i},
\end{align}
and
\begin{align}
  W  &= P_0-i\Tilde{P}_0,\nonumber\\
  q_i&=ib_i+ b_jb_k,\,\, i\neq j\neq k,
\end{align}
with 
\begin{align}
  P_0  &= 1+b_1^2+b_2^2+b_3^2,\nonumber\\
  \Tilde{P}_0 &=2b_1b_2b_3.
\end{align}

In regard to the potential, we consider the parent supergravity $N = 8$ to carry either of the subsequent gauge groups: $\textsc{SO(8)}$ \cite{deWit:1981sst}, its CSO contraction $\textsc{SO(8)}\ltimes\mathbb{R}^{12}$ \cite{Cordaro:1998tx, Hull:1984qz} or the dyonic CSO $[\textsc{SO(6)}\times\textsc{SO(2)}]\ltimes\mathbb{R}^{12}$ gaugings \cite{DallAgata:2011aa, DallAgata:2014tph}, which all have a higher-dimensional interpretation. These gaugings may be qualified by the embedding of tensors having non-vanishing components in the $\mathbf{36}'\oplus\mathbf{36}$ of $\textsc{SL}(8,\mathbb{R})$. The components $ \theta_{AB}$ and $\xi_{AB}$ are given by \cite{Larios:2023lxg}
\begin{align}
    \theta&=g\,\textrm{diag}\left(1,1,1,1,1,1,x,x\right),\\
    \xi&=m'\,\textrm{diag}\left(0,0,0,0,0,0,\Tilde{x},\Tilde{x}\right),
\end{align}
or as mentioned, $\Tilde{x}$ and $x$ are non-vanishing components in the $\mathbf{36}'\oplus\mathbf{36}$. Here, $g$ and $m'$ are the electric and magnetic coupling constants, respectively. For the sake of simplicity, these are valid gauges fulfilling linear and quadratic constraints for the embedding tensor as long as $x\Tilde{x}= 0$ \cite{Inverso:2012hs}. The three nonequivalent candidates are \cite{Larios:2023lxg}
\begin{align}
    \theta_{(8)} &=g\,\textrm{diag}\left(1,1,1,1,1,1,1,1\right),\,\, \xi_{(8)}=0,\label{x}\\
     \theta_{(6)} &=g\,\textrm{diag}\left(1,1,1,1,1,1,0,0\right),\,\, \xi_{(6)}=0,\label{y}\\
\end{align}
and
\begin{align}
    \theta_{(6c)} &=g\,\textrm{diag}\left(1,1,1,1,1,1,0,0\right),\nonumber\\
    \xi_{(6c)} &=m'\,\textrm{diag}\left(0,0,0,0,0,0,1,1\right),\label{z}
\end{align}
with the labels designating the three different gauge groups above, respectively, and $c =m'/g\neq 0$ in the latter case. 

Within the $\mathbb{S}\mathbb{T}\mathbb{U}$ truncation, these \textcolor{black}{embedding} tensors induce Fayet-Iliopoulos \textcolor{black}{gaugings}, whose potentials are defined as follows \cite{Larios:2023lxg}
\begin{align}
    V_{(8)}&=-4g^2\sum_i\left(\Tilde{Y}_i^2+Y_i^2\right),\label{v1}\\
V_{(6)}&=V_{(6c)}=-4g^2\left(\Tilde{Y}_1^2+Y_2^2+Y_3^2\right)\label{v2}.
\end{align}
The potential \eqref{v1} exhibits just one critical point. It is located at the scalar origin and maps onto the maximally supersymmetric solution $\textsc{SO(8)}$. The potentials \eqref{v2}, on the other hand, lack an extremum in this sector.

Although the $V_{(6c)}$ potential is unaware of the value of the magnetic coupling $m'$, fermion couplings in this theory rely on it. A comparable scenario has been formerly studied in the $\mathbb{S}\mathbb{T}\mathbb{U}$ truncation of dyonic-gauged $\textsc{SO(8)}$ supergravity \cite{Lu:2014fpa}. As a matter of fact, the truncated theory is not supersymmetric for non-vanishing $m'$. The electric cases, on the other hand, are gauged $N = 2$ supergravities coupled to three vector multiplets. To confirm this, it should be noted that theories with $N = 2$ supersymmetries can necessarily be recovered from the canonical viewpoint of \cite{Trigiante:2016mnt, Andrianopoli:1996cm} in terms of special K\"{a}hler and quaternionic structures. For the scalar manifold \eqref{sm}, the special holomorphic section may be defined as \cite{Larios:2023lxg}
\begin{equation}
    \Omega^M(z)=\lbrace1,u_1,u_2,u_3,-u_1u_2u_3,u_2u_3,u_1u_3,u_1u_2\rbrace\,,
\end{equation}
in standard holomorphic coordinates in \eqref{sm}. This section outlines the geometry of the scalar manifold and encrypts its K\"{a}hler potential using the following equation \cite{Larios:2023lxg}:
\begin{widetext}
    \begin{equation}
    \mathcal{K}=-log[i\Bar{\Omega}^M\mathbb{C}_{MN}\Omega^N]=-log[-i (z_1-\Bar{z}_1) (z_2-\Bar{z}_2)(z_3-\Bar{z}_3)],
\end{equation}
\end{widetext}
where $z_i$ and $\Bar{z}_i$ are complex and conjugate complex scalar fields and $\mathbb{C}_{MN}$ being the symplectic form on $\textsc{Sp(8},\mathbb{R})$. Purely Fayet-Iliopoulos gaugings have a potential \cite{Trigiante:2016mnt}
\textcolor{black}
{\begin{equation}
    V=4g^2\left(g^{i\Bar{i}}D_i V^M D_{\Bar{i}}\Bar{V}^N-3V^M\Bar{V}^N\right)\upsilon_M\upsilon_N\,,
\end{equation}
with covariantly holomorphic sections
\begin{equation}
\label{covder}
    D_{\Bar{i}}V^M=\left(\partial_{\Bar{i}}-\frac{1}{2}\partial_{\Bar{i}}\mathcal{K}\right)V^M=0.
\end{equation}
Here, $D$ denotes the derivative covariantized with respect to the K\"{a}hler  bundle $U(1)$, under which $V^M$ has non-vanishing holomorphic weight~\cite{Bellucci2,Trigiante:2016mnt}}, $g_{i\Bar{i}}=\partial_i\partial_{\Bar{i}}\mathcal{K}$ is the Hermite metric associated with the K\"{a}hler potential, $\upsilon_M$ is the integration tensor depicting how the gauge group $\textsc{U(1)}$ is embedded in the R-symmetry group $\textsc{SU(2)}$, and  $V_M$ a section of the special bundle $\textsc{U(1)}$. {\color{black}In fact, $V_M$ could be described as a $8$-component vector of complex functions $V^M=V(z,\Bar{z})$:
\begin{equation}
    V(z,\Bar{z})=\begin{pmatrix}
        L^\Lambda(z,\Bar{z})\\
         M^\Lambda(z,\Bar{z})
    \end{pmatrix},\qquad \Lambda=0,\cdots,3.
\end{equation}
  It is worth mentioning that $V^M$ satisfies the symplectic constraint $i\braket{V,\Bar{V}}=1$, as well as being related to the holomorphic symplectic vector $(X^\Lambda, F_\Lambda)^T$ by
\begin{equation}
    V^M=e^{\mathcal{K}/2}  \Omega^M(z)=e^{\mathcal{K}/2}\begin{pmatrix}
        X^\Lambda(z)\\
         F^\Lambda(z)
    \end{pmatrix}.
\end{equation}
So, the symplectic vector $V^M$ for such a prepotential is given by
\begin{equation}
    V^M=e^{\mathcal{K}/2}  (1,u_1,u_2,u_3,-u_1u_2u_3,u_2u_3,u_1u_3,u_1u_2)^T.
\end{equation}
where the symplectic sections are given explicitly by 
\begin{align}
    L^\Lambda&=e^{\mathcal{K}/2}(1, u_1, u_2, u_3)^T,\\
       M^\Lambda&=e^{\mathcal{K}/2}(-u_1u_2u_3, u_2u_3, u_1u_3, u_1u_2)^T.
\end{align}
However, the consideration of an axion-free case in this model plays a crucial role in unveiling such limits of symplectic structure. Thus, parametrize the purely imaginary scalar fields as $u_i =-i e^{-\phi_i} =-i\lambda_i$, with $\lambda^i$ real and positive. Along with this condition, one has
\begin{align}
    Y_i= \Tilde{Y}_i^{-1}=e^{\phi_i}=\lambda_i^{-1},\\ (Y_i \Tilde{Y}_i)^2=1,\\ b_i=q_j=0,\\ W=1,
\end{align}
while, the K\"{a}hler potential can be given as
\begin{equation}
    \mathcal{K}=-log(8\lambda_1\lambda_2\lambda_3),
\end{equation}
so that 
\begin{equation}
e^{\mathcal{K}/2}=1/(2\sqrt{2}\sqrt{\lambda_1\lambda_2\lambda_3}),
\end{equation}
 which requires to be on the branch of positive $\lambda_i$'s.

 The symplectic sections for our configuration are
 \begin{widetext}
     \begin{equation}
    L^\Lambda=\frac{1}{2\sqrt{2}}\begin{pmatrix}
       1/\sqrt{\lambda_1\lambda_2\lambda_3}\\-i\sqrt{\lambda_1/\lambda_2\lambda_3}\\-i\sqrt{\lambda_2/\lambda_1\lambda_3}\\-i\sqrt{\lambda_3/\lambda_1\lambda_2}
    \end{pmatrix},\qquad M^\Lambda=\frac{1}{2\sqrt{2}}\begin{pmatrix}
       -i\sqrt{\lambda_1\lambda_2\lambda_3}\\-i\sqrt{\lambda_2\lambda_3/\lambda_1}\\-i\sqrt{\lambda_1\lambda_3/\lambda_2}\\-i\sqrt{\lambda_1\lambda_2/\lambda_3}
    \end{pmatrix},
\end{equation}
 \end{widetext}
 where the period matrix in question of the axion-free case can be given as follows
    \begin{equation}\mathcal{N}_{ab}=i
    \begin{pmatrix}
-\lambda_1\lambda_2\lambda_3 &0  &  0&0 \\
 0& -\frac{\lambda_1}{\lambda_2\lambda_3} &  0 & 0\\
0 & 0 & -\frac{\lambda_2}{\lambda_1\lambda_3} &0 \\
0 & 0 &  0& -\frac{\lambda_3}{\lambda_1\lambda_2}
\end{pmatrix}.
\end{equation}
}

In this language, the potential obtained from the gauging truncation $\textsc{SO(8)}$ is defined by \cite{Larios:2023lxg}
\begin{equation}
    \upsilon_M^{(8)}=g\left(-1,0,0,0,0,1,1,1\right),
\end{equation}
while the embedding tensor associated with $V_{(6)}$ is yielded by 
\begin{equation}
    \upsilon_M^{(6)}=g\left(-1,0,0,0,0,0,1,1\right).
\end{equation}
For electric theories, the $N = 2$ supersymmetry variations and fermionic mass terms in the Lagrangian associated with these embedding tensors are in agreement with the $N = 8$ truncation of fermion shifts related to Eqs. \eqref{x}-\eqref{z} \cite{Larios:2023lxg}. Notwithstanding, for the dyonic gauging in Eqs. \eqref{x}-\eqref{z}, the fermion shifts rely on $m$, which means that they remain unrecoverable in the $N = 2$ languages.

Alternatively, the potentials in \eqref{v1} could similarly be generated from $V_{(8)}$ applying singular scaling. This requires the scalars to be redefined as 
\begin{align}
    \phi_1&\rightarrow\phi_1-k,\,\,\,\,\phi_{2,3} \rightarrow\phi_{2,3} +k,\nonumber\\
     \chi_1&\rightarrow e^k\chi_1,\,\,\,\,\,\,\,\,\,\chi_{2,3} \rightarrow e^{-k} \chi_{2,3}.
\end{align}
Likewise, the gauge coupling can be scaled as $g\rightarrow e^{-k/2}g$, while the gauge fields, namely the three fields of the vector multiplet and that of the supergravity multiplet, i.e., the graviphoton, must be scaled as
\begin{equation}
\label{scaling}
    A_{1,2,3} \rightarrow e^{k/2}A_{1,2,3},\,\,\,\,\,\,\,\,\,A_{4} \rightarrow e^{-3k/2} A_{4}\,,
\end{equation}
whereas the metric remains invariant. It should be noted that 
the graviphotonic index, usually denoted by "0", is here indicated by "4". Consistent with our choice of the "4D/5D special coordinates" symplectic frame, the scaling~\eqref{scaling} is precisely the $SO(1,1)$ dilatational transformation acting on the Maxwell potentials, where $SO(1,1)$ is the Kaluza-Klein one, associated to the $S^1$ radius of the $D=5\rightarrow D=4$ Kaluza-Klein dimensional reduction.
Moreover, the singular limit $k\rightarrow\infty$ on \eqref{lg} subsequent to the redefinitions~\eqref{scaling} brings the Lagrangian $\mathcal{L}_{(8)}$ to $\mathcal{L}_{(6)}$, also involving fermionic coupling. \textcolor{black}{Clearly, this limit would have the effect to suppress the graviphoton Maxwell field (i.e., $A_4=0$).}

{\color{black}At this point, after recalling the basic \textcolor{black}{facts} of the $\mathbb{S}\mathbb{T}\mathbb{U}$ supergravity model} {\color{black} and proceeding with the exploration of the AdS-BH limit, it is worth making the following observation: in the realm of $\mathbb{S}\mathbb{T}\mathbb{U}$ supergravity, the $\textsc{SO}(8)$-truncated symplectic basis exhibits the peculiar feature that the FI potential $V_{F I,\textsc{SO}(8)}$ admits critical points at a finite distance from the origin~\cite{Bellucci2}. Conversely, within the $\textsc{SO}(1, 1)^2$ and $\textsc{SO}(2, 2)$ covariant bases, the $V_{F I, \textsc{SO}(1,1)^2}$ and $V_{F I, \textsc{SO}(2,2)}$ potentials exhibit only runaway behavior, admitting no more than minima at infinite distance from the origin~\cite{Bellucci2}.} 

\textcolor{black}{On the other hand, with the application of the embedding tensor and tensor hierarchies as tools, the embedding of any solution of the four-dimensional gauged $\mathbb{S}\mathbb{T}\mathbb{U}$  models into M-theory or type IIB supergravity can be achieved by following~\cite{Larios:2023lxg}. Furthermore, to expand on our knowledge of this mechanism, another example is the massive type IIA supergravity \cite{Romans:1985tz}, which admits a consistent truncation on the six-sphere $S^6$ to maximal, $N = 8$, supergravity in four dimensions. The relevant $D = 4$ gauge group is the non-simple $\textsc{ISO}(7) = \textsc{SO}(7) \ltimes \mathbb{R}^7$, and the gauging is of the dyonic type extensively reported in \cite{DallAgata:2012mfj,DallAgata:2014tph}. In contrast to the $N = 8$ $D = 4$ $\textsc{SO}(8)$ and $D = 5$ $\textsc{SO}(6)$  gaugings, the $\textsc{ISO}(7)$ dyonic gauging does not admit an $AdS$ $N = 8$ vacuum that could eventually rise to a maximally supersymmetric $AdS_4\times S^6$ Freund-Rubin background of massive type IIA. }

\subsection{AdS-BH limits}
\textcolor{black}{For the purpose of our next analysis, we refer to~\cite{Larios:2023lxg}.
In this regard,  we would like to point out that a change of symplectic frame is involved here, allowing for four-charge $AdS_4$ (non-extremal) black hole solutions with Fayet-Iliopoulos $\textsc{U}(1) \in \textsc{SU}(2)_R$ Abelian gaugings of $N=2$, $D=4$ supergravity (see~\cite{Bellucci2,Larios:2023lxg} for more details). For the extremal limit of these solutions, one may refer to~\cite{Bellucci2}.}

By freezing the axions, the potential \eqref{v1} in the $\mathbb{S}\mathbb{T}\mathbb{U}$ truncation of the $\textsc{SO(8)}$ gauge takes the form
\begin{equation}
V=-8g^2\left(\cosh{\phi_1}+\cosh{\phi_2}+\cosh{\phi_3}\right).
\end{equation}\label{fi}
 The $\mathbb{S}\mathbb{T}\mathbb{U}$ theory admits the four-charge AdS-black hole solutions \cite{27,Cvetic:1999xp,Larios:2023lxg} }
\begin{align}
\textrm{d}s^2 &= -\prod_{i=1}^4H_i^{-1/2}f(r)\textrm{d}t^2
+\prod_{i=1}^4H_i^{1/2}\left(f(r)^{-1}\textrm{d}r^2+r^2\textrm{d}\Omega^2\right),\label{metric}\nonumber\\
A^i&=\frac{\sqrt{q_i(q_i+2m)}}{r+q_i}dt,
\end{align}
{\color{black}
with
\begin{align}e^{2\phi_1}&=\frac{H_1H_2}{H_3H_4},\,\,e^{2\phi_2}=\frac{H_1H_3}{H_2H_4},\,\,e^{2\phi_3}=\frac{H_1H_4}{H_2H_3},\nonumber
\label{metric}
\end{align}}
and
\begin{equation}
f(r)=1-\frac{2m}{r}+g^2r^2\prod_{i=1}^4H_i, \   \  \  \ H_i=1+\frac{q_i}{r}.
\label{Eq6}
\end{equation}
{\color{black}Here $m$ and $q_i$ are the bare mass and charge parameters related to the ADM mass and charge of black holes, respectively}. 
{\color{black}The ADM mass $M$ and the charge $Q$ of black holes are calculated by using the conserved charges, leading to \cite{Lu:2013eoa}}
\begin{eqnarray}
M&=&m+\frac{1}{4}\sum_iq_i,\label{Eq7}\\ 
Q_i&=&\frac{1}{2}\sqrt{q_i(q_i+2m)}. \label{Eq8}
\end{eqnarray}
{\color{black}Furthermore,}  the set of thermodynamic quantities of the $\mathbb{S}\mathbb{T}\mathbb{U}$ black hole is expressed as follows:
\begin{eqnarray}
S&=&\pi\prod_i\sqrt{\left(r_h+q_i\right)}, \label{Eq9}\\
T&=&\frac{f'(r_h)}{4\pi}\prod_iH_i^{-1/2},\label{Eq10}\\ 
\Phi^i&=&\frac{\sqrt{q_i(q_i+2m)}}{2\left(r_h+q_i\right)},\label{Eq11}\\ 
V&=&\frac{\pi}{3}r_h^3\prod_iH_i\sum_j\frac{1}{H_j},\label{Eq12}
\end{eqnarray}
where the above-mentioned quantities denote, respectively,  the entropy, the Hawking temperature, the electrostatic potential at infinity, and the thermodynamic volume. 

In addition, the gauge coupling constant and the AdS radius behave like the thermodynamic pressure according to the following expression \cite{A,B,TranNHung:2024pig}:
\begin{equation}
    P=\frac{3\,g^2}{8\pi}=\frac{3}{8\,\pi\ell^2}\,,
    \label{P}
\end{equation}
where $\ell$ is the AdS radius.

{\color{black}\subsection{ Horizon properties}
\label{HP}
In this section we discuss the properties of the horizon structure of the R-charged black hole with four abelian gauge fields $\textsc{U}(1)^4$, i.e., $(q_1, q_2, q_3,q_4)$. The key properties for this purpose are examined by assuming, at the horizon radius $r_h$, the constraint $g^{rr}= 0$, which is clearly expressed in terms of
\begin{widetext}
    \begin{equation}
    f(r_h)=1-\frac{m}{r_h^2}+ g^2 r_h^2 \bigg(1+\frac{q_1}{r_h^2}\bigg) \bigg(1+\frac{q_2}{r_h^2}\bigg)
   \bigg(1+\frac{q_3}{r_h^2}\bigg)\bigg(1+\frac{q_4}{r_h^2}\bigg)= 0.
\end{equation}
\end{widetext}
Concretely, $f(r)$ has a set of roots such that $r = r_h$; ergo, $f(r_h) = 0$. This condition can be expressed more clearly as
    \begin{eqnarray}\label{horizon}
    g^2 \prod_{i= 1}^4 \left(r_h+q_i\right) +r_h \left(r_h-2 m\right)&=&0.
\end{eqnarray}
The analysis concerning the horizon structure of the $\mathbb{S}\mathbb{T}\mathbb{U}$ black hole is done by examining the solution set of Eq. \eqref{horizon}. In the following computations, we take into account four charge configurations on the parameter space of the abelian gauge fields $\mathcal{M}_q(q_1,q_2,q,_3,q_4)$, namely:  
\begin{itemize}
\item     case 1: $q_1=q_2=q_3=q_4=q$.  In this case, when $m>0$, Eq. \eqref{horizon} has two distinct physical solutions, namely, the inner and outer horizon radii, as shown in Fig. \ref{F}. Such solutions are explicitly given by:
\begin{widetext}
\begin{align}
    r_{1,2}&=\frac{1}{6} \Biggl\{\sqrt{3} \Bigg(\frac{1}{g^2}\bigg(\frac{12 \sqrt{3} (m+q)}{A}\pm\bigg(18 g^2 \left(3 m^2+2 m q+q^2\right)+6 \sqrt{3} \bigg(g^2 \bigg(-16 g^4 q^3 (2 m+q)^3+g^2
   \bigg(27 m^4\nonumber\\
   &+36 m^3 q+14 m^2 q^2-4 m q^3-q^4\bigg)+m^2\bigg)\bigg)^{1/2}+1\bigg)^{1/3}+\frac{-12 g^2 q (2 m+q)-1}{\sqrt[3]{B}}-4\bigg)\Bigg)^{1/2} +\sqrt{3} A-6 q\Biggr\}, 
\end{align}
\end{widetext}
where
\begin{widetext}
    \begin{align}
         A &=\sqrt{\frac{12 g^2 q (2 m+q)+\left(\sqrt[3]{B}-1\right)^2}{g^2 \sqrt[3]{B}}},\nonumber\\
   B &=18 g^2 \left(3 m^2+2 m q+q^2\right)+6 \sqrt{3} \sqrt{g^2
   \left(-16 g^4 q^3 (2 m+q)^3+g^2 \left(27 m^4+36 m^3 q+14 m^2 q^2-4 m q^3-q^4\right)+m^2\right)}+1.\nonumber
    \end{align}
\end{widetext}
On the other hand, the physical set of the horizon radius tends to be empty as long as the bare mass parameter is reduced to zero.
\item Case 2: $q_1=q_2=q_3=q$\, and $q_4=0$.
    \textcolor{black}{In this case, which corresponds to unfluxing the $N=2, D=4$ graviphoton,} Eq.~\eqref{horizon} provides only one physical horizon radius. This always exists as long as the value of the bare mass parameter is a positive definite. Graphically, this fact is shown in Fig.~$\ref{F}$.  Analytically, the solution is given by:
    \begin{widetext}
        \begin{equation}
r_{1}  = \frac{\sqrt[3]{9 g^4 (2 m+q)+\sqrt{3} \sqrt{g^6 \left(27 g^2 (2 m+q)^2+4\right)}}}{\sqrt[3]{2} 3^{2/3} g^2}-\frac{\sqrt[3]{\frac{2}{3}}}{\sqrt[3]{9 g^4 (2
   m+q)+\sqrt{3} \sqrt{g^6 \left(27 g^2 (2 m+q)^2+4\right)}}}-q.\label{s1}
\end{equation}
    \end{widetext}
\item Case 3: $q_1=q_2=q$\, and $q_3=q_4=0$. 
\textcolor{black}{This configuration corresponds to unfluxing the graviphoton and one vector multiplet's Maxwell field.}
   As for the case 2, in this setting Eq. \eqref{horizon} provides only one physical horizon radius, which always exists as long as the value of the bare mass parameter is a positive definite. Graphically, this is shown in Fig.~$\ref{F}$, corresponding to the analytical solution:
    \begin{widetext}
        \begin{align}
r_{1}  =  \frac{1}{3} \Biggl\{-2q&+\frac{\sqrt[3]{g^6 q^3+9 g^4 (3 m+q)+3 \sqrt{3} \sqrt{g^{10} q^3 (2 m+q)+g^8 \left(27 m^2+18 m q+2 q^2\right)+g^6}}}{g^2}\nonumber\\
&+\frac{g^2
   q^2-3}{\sqrt[3]{g^6 q^3+9 g^4 (3 m+q)+3 \sqrt{3} \sqrt{g^{10} q^3 (2 m+q)+g^8 \left(27 m^2+18 m q+2 q^2\right)+g^6}}}\Biggr\}.\label{s2}
\end{align}
    \end{widetext}

\item Case 4: $q_1=q$,\, and $q_2=q_3=q_4=0$. 
\textcolor{black}{This case corresponds to unfluxing the graviphoton and two vector multiplets' Maxwell fields.}
Again,  Eq. \eqref{horizon} provides only one physical horizon radius, which always exists as long as the value of the bare mass parameter is a positive definite. Graphically, this fact is shown in Fig.~$\ref{F}$, which is associated with the solution
    \begin{widetext}
        \begin{align}
r_{1}  = \frac{1}{6} \Biggl\{-2q&+\frac{2^{2/3} \sqrt[3]{-2 g^6 q^3+9 g^4 (6 m+q)+3 \sqrt{3} \sqrt{g^6 \left(-8 g^4 m q^3+g^2 \left(108 m^2+36 m q-q^2\right)+4\right)}}}{g^2}\nonumber\\
&+\frac{2
   \sqrt[3]{2} \left(g^2 q^2-3\right)}{\sqrt[3]{-2 g^6 q^3+9 g^4 (6 m+q)+3 \sqrt{3} \sqrt{g^6 \left(-8 g^4 m q^3+g^2 \left(108 m^2+36 m q-q^2\right)+4\right)}}}\biggr\}.\label{s3}
\end{align}
    \end{widetext}
\end{itemize}
 \begin{figure*}[tbh!]
    \centering
    \begin{subfigure}[b]{0.5\textwidth}
        \centering
        \includegraphics[scale=0.79]{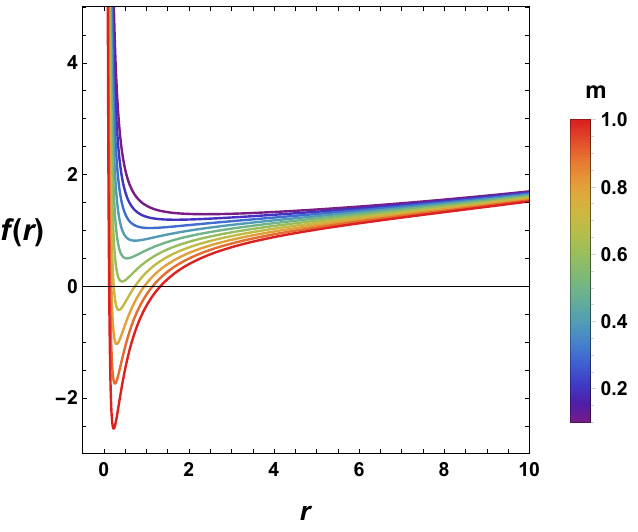}
        \caption{$q_1=q_2=q_3=q_4=q$.}
        \label{subfig:wec1}
    \end{subfigure}%
    \hfill
    \begin{subfigure}[b]{0.5\textwidth}
        \centering
        \includegraphics[scale=0.79]{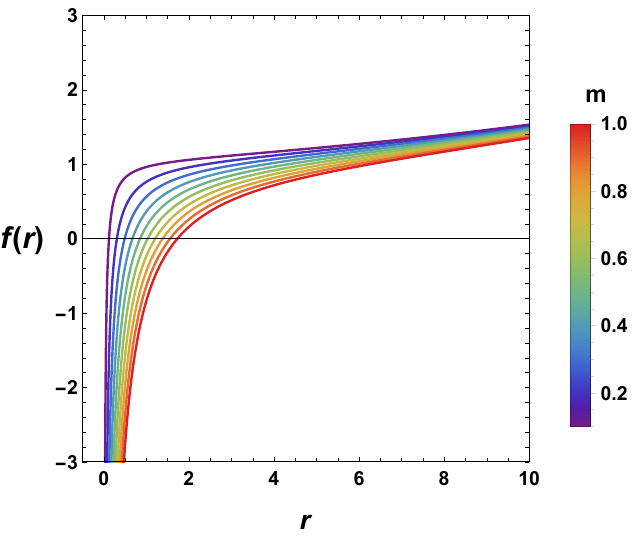}
        \caption{$q_1=q_2=q_3=q$ and $q_4=0$.}
        \label{subfig:nec1}
    \end{subfigure}%
    \\
    \begin{subfigure}[b]{0.5\textwidth}
        \centering
        \includegraphics[scale=0.79]{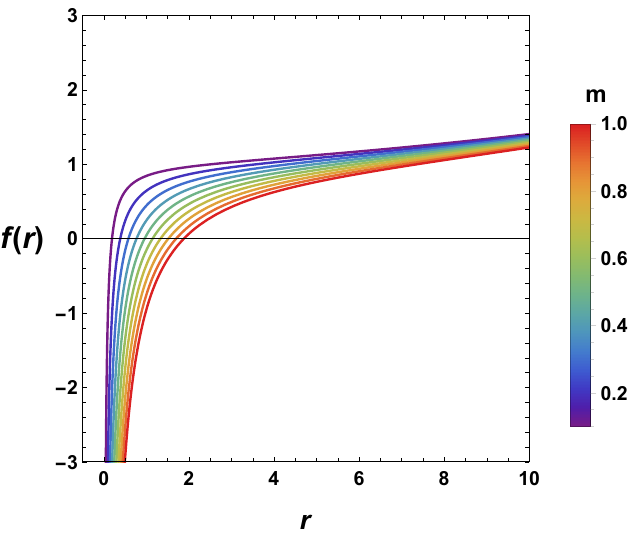}
        \caption{$q_1=q_2=q$ and $q_3=q_4=0$.}
        \label{subfig:wec2}
    \end{subfigure}%
    \hfill
    \begin{subfigure}[b]{0.5\textwidth}
        \centering
        \includegraphics[scale=0.79]{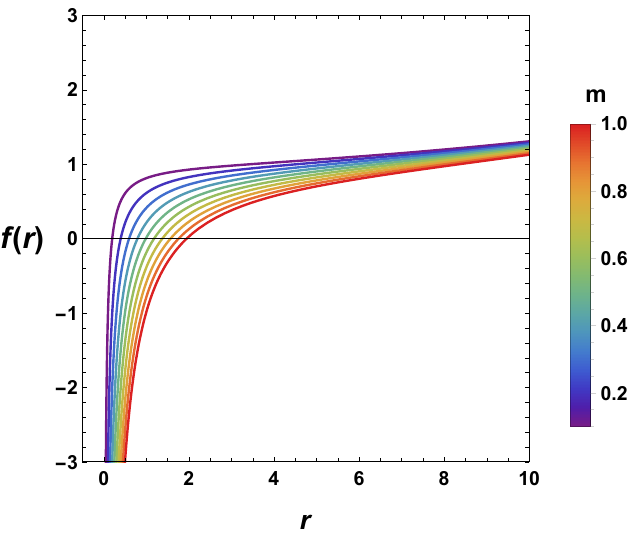}
        \caption{$q_1=q$ and $q_2=q_3=q_4=0$.}
        \label{subfig:nec2}
    \end{subfigure}%
    \caption{{\color{black} Plot of $f(r)$ versus $r$, for different values of $m$ and four charge configurations. We set $g=0.05$.}}
    \label{F}
\end{figure*}
 A simple observation shows that only the case with four equal charges admits two horizon radii, i.e., event and Cauchy radii. While the other cases admit a single horizon radius that is analytically present in each case by Eqs. \eqref{s1}-\eqref{s3}. \textcolor{black}{Notice that a plausible explanation for this finding may be related to the fact that the case 1 is the only configuration having non-vanishing graviphotonic charge, and thus admitting a non-singular Reissner–Nordstr\"{o}m black hole limit. }
 }

\section{Optical Properties of the black hole}
\label{S3}

This section is devoted to study black hole optical properties and, particularly, shadow behaviors. 

\subsection{ Shadow behaviors}
Let us consider the null geodesic motion in $g_{\mu\nu}(x)$ background. The first step is to use the Lagrangian formalism \cite{44,45,46} as a useful tool to describe the geodesic process. The Lagrangian in the background $(\ref{metric})$ is given as:
\begin{widetext}
\begin{equation}
   \mathcal{L}=-\frac{f(r)}{2}\,\prod_{i=1}^4H_{i}^{-1/2}\,\dot{t}^2+\frac{f(r)^{-1}}{2}\,\prod_{i=1}^4H_{i}^{1/2}\,\dot{r}^2+\frac{r^2}{2}\,\prod_{i=1}^4H_{i}^{1/2}\mathcal{L}_{\Omega_{2}},
    \label{Lag}
\end{equation}
\end{widetext}
where $\mathcal{L}_{\Omega_{2}}$ is the angular Lagrangian, which can be written as
\begin{equation}
\mathcal{L}_{\Omega_{2}}=\dot{\theta}^2+\sin^2\theta\,\dot{\phi}^2.
    \label{E5-2}
\end{equation}
Here, the overdot denotes differentiation with respect to the affine parameter $\delta$ along the geodesic. Because the $\left(t,\phi\right)$ variables are cyclic, the underlying Lagrangian analysis is independent of them, and thus the associated conjugate momenta $\pi_q=\partial\mathcal{L}/\partial\dot{q}$ are conserved quantities. Specifically, we have
\begin{align}
    \pi_t&=-f(r)\prod_{i=1}^4 H_i^{-1/2}\,\dot{t}\equiv-E,\\
    \pi_\phi&=r^2\prod_{i=1}^4H_i^{1/2}\,\sin^2\theta\,\dot{\phi}\equiv L,
    \label{E5-2bis}
\end{align}
where $E$ is a positive constant measuring the temporal invariance of the Lagrangian. It should be noted that this constant has no connection to energy due to the fact that the metric system $(\ref{metric})$ is not asymptotically flat. On the other hand, the condition $L=const.$ ensures the conservation of the angular momentum.

A more convenient approach for describing analytically the motion of a particle around the $\mathbb{S}\mathbb{T}\mathbb{U}$ black hole is provided by the Hamilton Jacobi formalism along with the Carter approach \cite{17}. In this regard, the Hamilton-Jacobi equation is given in the following way:
\begin{eqnarray}
    \frac{\partial S}{\partial \delta}=-\frac{1}{2}g^{\mu\nu}\frac{\partial S}{\partial x^\mu}\frac{\partial S}{\partial x^\nu},
    \label{S}
\end{eqnarray}
where $S$ is the Jacobi action of the test particle and $\delta$ is the affine parameter as defined above. Incorporating the inverse metric component of the background $(\ref{metric})$ into the Eq.$(\ref{S})$, one can obtain
\begin{widetext}
    \begin{equation}
        -2\frac{\partial S}{\partial \delta}=-\frac{\prod_{i=1}^4H_i^{1/2}}{f(r)}\left(\frac{\partial S_t}{\partial t}\right)^2+\frac{f(r)}{\prod_{i=1}^4}\left(\frac{\partial S_r}{\partial r}\right)^2+\frac{1}{r^2\prod_{i=1}^4H_i^{1/2}}\left(\frac{\partial S_\theta}{\partial \theta}+\frac{1}{\sin^2\theta}\left(\frac{\partial S_\phi}{\partial \phi}\right)^2\right).
        \label{-2}
    \end{equation}
\end{widetext}
It is worth noting that a separable solution for the Jacobi action is explicitly represented by
\begin{equation}
    S=\frac{1}{2}\mu^2\delta-Et+L\phi+S_r(r)+S_\theta(\theta),
    \label{Jacobi}
\end{equation}
where $\mu$ is the rest mass of the test particle. In what follows, we restrict the study to massless particles such as photons $(\mu=0)$, which are supposed to {\color{black}generate null geodesic motions (photon orbits)}. As a result, implementing Jacobi's action $(\ref{Jacobi})$ on Eq. $(\ref{-2})$ leads to the following expression
\begin{widetext}
    \begin{align}
        0&=\bigg\lbrace\frac{\prod_{1=1}^4H_i^{1/2}}{f(r)}E^2-\frac{f(r)}{\prod_{i=1}^4H_i}\left(\frac{\partial S_r}{\partial r}\right)^2-\frac{1}{r^2\prod_{i=1}^4H_i^{1/2}}\left(L^2+\mathcal{C}\right)\bigg\rbrace\nonumber\\
        &-\bigg\lbrace\frac{1}{r^2\prod_{i=1}^4H_i^{1/2}}\left(\frac{\partial S_\theta}{\partial\theta}\right)^2+\frac{1}{r^2\prod_{i=1}^4H_i^{1/2}}\left(L^2\cot^2\theta-\mathcal{C}\right)\bigg\rbrace,
    \end{align}
\end{widetext}
where $\mathcal{C}$ is the Carter constant. After executing some computations, we are led to the following independent couple of equations:
\begin{align}
    r^4f^2(r)\left(\frac{\partial S_r}{\partial r}\right)^2&=r^4\prod_{i=1}^4H_i\,E^2-r^2\left(L^2+\mathcal{C}\right)f(r),\\
    \left(\frac{\partial S_\theta}{\partial \theta}\right)^2&=\mathcal{C}-L^2\cot^2\theta.
\end{align}
In accordance with the Hamilton-Jacobi formalism, the appropriate set of the geodesic equation is given by
\begin{align}
\frac{dt}{d\delta}&=\frac{\prod_{i=1}^4H_i^{1/2}}{f(r)}E,\label{1}\\
r^2\prod_{i=1}^4H_i^{1/2}\frac{dr}{d\delta}&=\pm\sqrt{\mathcal{R}},\label{2}\\
r^2\prod_{i=1}^4H_i^{1/2}\frac{d\theta}{d\delta}&=\pm\sqrt{\Theta},\label{3}\\
\frac{d\phi}{d\delta}&=\frac{L}{r^2\prod_{i=1}^4H_i^{1/2}\sin^2\theta},\label{4}
\end{align}
where the signs $"+/-"$ indicate the radial directions in which photons are moving - outward and inward, respectively. To complete the analysis, the $\mathcal{R}$ and $\Theta$ are expressed as follows:
\begin{align}
    \mathcal{R}&=r^4\prod_{i=1}^4H_i\, E^2-r^2\left(L^2+\mathcal{C}\right)f(r),\\
    \Theta&=\mathcal{C}-L^2\cot^2\theta.
\end{align}
The photon's motion in the space-time $(\ref{metric})$ is controlled by Eqs. $(\ref{1})$-$(\ref{4})$.

\begin{figure*}[tbh!]
    \centering
    \begin{subfigure}[b]{0.5\textwidth}
        \centering
        \includegraphics[scale=0.79]{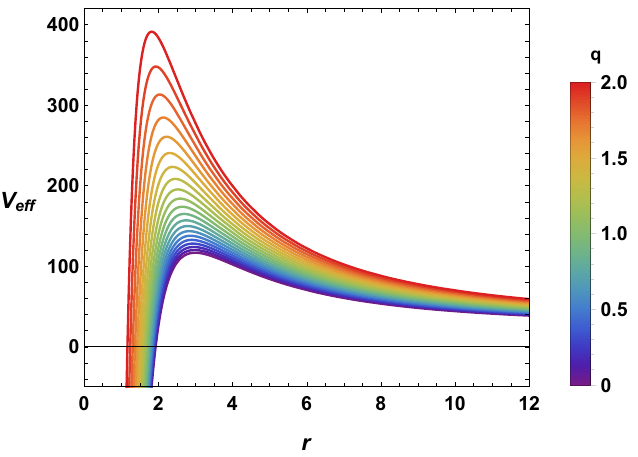}
        \caption{$q_1=q_2=q_3=q_4=q$.}
        \label{subfig:wec3}
    \end{subfigure}%
    \hfill
    \begin{subfigure}[b]{0.5\textwidth}
        \centering
        \includegraphics[scale=0.79]{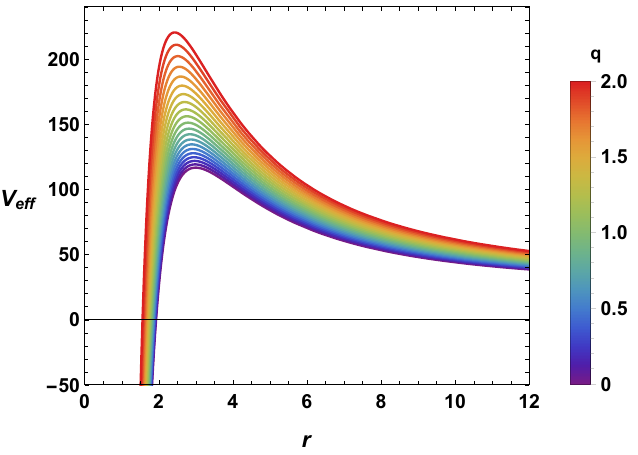}
        \caption{$q_1=q_2=q_3=q$ and $q_4=0$.}
        \label{subfig:nec3}
    \end{subfigure}%
    \\
    \begin{subfigure}[b]{0.5\textwidth}
        \centering
        \includegraphics[scale=0.79]{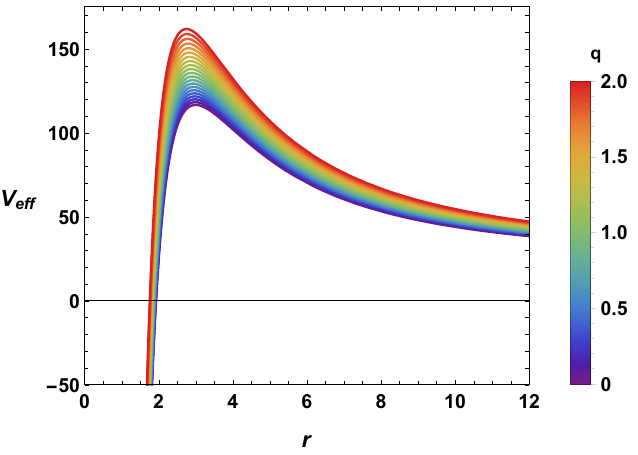}
        \caption{$q_1=q_2=q$ and $q_3=q_4=0$.}
        \label{subfig:wec4}
    \end{subfigure}%
    \hfill
    \begin{subfigure}[b]{0.5\textwidth}
        \centering
        \includegraphics[scale=0.79]{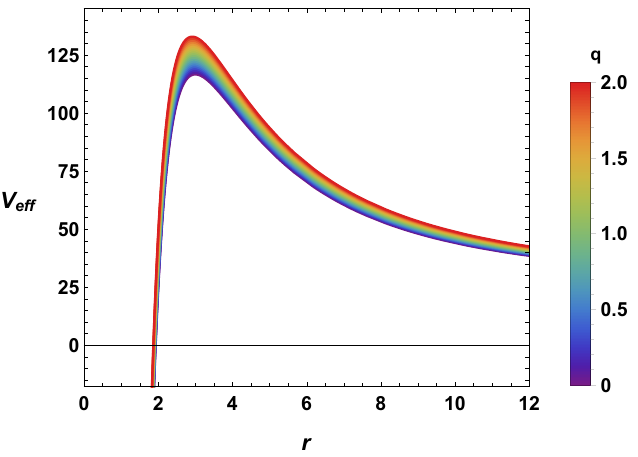}
        \caption{$q_1=q$ and $q_2=q_3=q_4=0$.}
        \label{subfig:nec4}
    \end{subfigure}%
            \caption{The graph of the radial evolution of the effective potential for four cases of charge configurations with $m=1$ and $g=0.1$.}
            \label{Fig1}
        \end{figure*}

To analyze in depth the shadow behaviors, it is required to define an effective potential. This step helps to show the shape of a black hole, which is entirely defined by the boundaries of its shadow and represents the apparent shape of the photon's unstable circular orbits. Thus, the effective potential is expressed in such a way that
\begin{equation}
    \left(\frac{dr}{d\delta}\right)^2+V_{eff}(r)=0.
    \label{E5-5}
\end{equation}
In other terms, the previous expression provides the equivalent of the following:
\begin{equation}
    V_{eff}(r)=f(r)\left\lbrace\frac{1}{r^2\prod_{i=1}^4H_i}\left(L^2+\mathcal{C}\right)-\frac{E^2}{f(r)}\right\rbrace.
    \label{E5-6}
\end{equation}
Fig.~$\ref{Fig1}$ shows the effective potential versus $r$ for the four charge configurations introduced in Sec.~\ref{HP}. In all cases, it is observed  that the effective potential increases as the charge parameter increases. 

We now remark that the maximum of the effective potential provides the needed information to
find the unstable circular orbit of the photons that carry information about the boundary of the apparent shape of the black hole. Therefore, we impose the following constraints \cite{Bar}:
\begin{equation}
    V_{eff}=\frac{dV_{eff}(r)}{dr}\biggr\rvert_{r=r_{ph}}=0,\quad 
     \mathcal{R}=\frac{d\mathcal{R}}{dr}\biggr\rvert_{r=r_{ph} }=0.
    \label{cons}
\end{equation}
From this, the photon sphere radius $r_{ph}$ can be related to the maximum of the effective potential and can therefore be explicitly deduced by solving the following equation:
\begin{widetext}
\begin{eqnarray}
\prod_{i=1}^4\left(1+\frac{q_i}{r_{ph}}\right)\left(r_{ph}f'(r_{ph})-2f(r_{ph})\right)-r_{ph}f({r_{ph}})\left(\prod_{i=1}^4H_i\right)'=0.
\label{E5-8}
\end{eqnarray}
\end{widetext}
In terms of the parameter space of the black hole system, one has
\begin{align}
  &2\left(r_{ph}-m\right)\prod_{i=1}^4q_i+4m\,r_{ph}^3\sum_{i=1}^4q_i\nonumber\\
  &-\left(\sum_{i=1}^4q_i-6m\right)r_{ph}^4
  -2r_{ph}^5+\mathcal{U}r_{ph}^2=0,
    \label{rph}
\end{align}
where
\begin{align}
\mathcal{U}&=q_1q_2q_3+q_2q_3q_4+q_1q_3q_4+q_1q_2q_4\nonumber\\
&+2\bigg\lbrace(q_3q_4+q_2(q_3+q_4)+q_1(q_2+q_3+q_4)\bigg\rbrace m.\nonumber
\end{align}

The steps for solving Eq. $(\ref{rph})$ are taken into consideration by choosing four cases of charge configurations as shown in Tab. $\ref{Tab1}$.
\begin{table*}[ht]
    \centering
    \begin{tabular}{clllll}
\hline
\hline
{case} &    Case 1 & Case 2 & Case 3 & Case 4 \\
\hline
$\mathcal{M}_q\left(q_1,q_2,q_3,q_4\right)$    &       $\mathcal{M}_q\left(q,q,q,q\right)$ &       $\mathcal{M}_q\left(q,q,q,0\right)$  & $\mathcal{M}_q\left(q,q,0,0\right)$ & $\mathcal{M}_q\left(q,0,0,0\right)$ \\
\hline
\end{tabular}
\caption{The four cases of the charge configurations.}
    \label{Tab1}
\end{table*}
Consequently, with respect to the four considered cases, a suitable solution for the photon radius $r_{ph}$ is expressed in terms of the bare mass parameter $\left(m\right)$ and the charge parameter $\left(q\right)$. Thus, $r_{ph}$ is explicitly specified by
\begin{widetext}
    \begin{equation}
r_{ph}= \begin{cases}
     \frac{1}{2} \left(\sqrt{9 m^2+2 m q+q^2}+3 m+q\right),\,\,\,\,\,\,\text{for}\,\, q_1=q_2=q_3=q_4=q, \vspace{3mm}\\
    \frac{1}{2} (6 m+q),\,\,\qquad\qquad\qquad\qquad\qquad\,\,\,\,\text{for}\,\,q_1=q_2=q_3=q\, \text{and}\,\, q_4=0,\vspace{3mm}\\
     \frac{1}{2} \left(\sqrt{m} \sqrt{9 m+4 q}+3 m\right),\,\,\qquad\quad\,\,\,\,\,\,\,\,\text{for}\,\,q_1=q_2=q\, \text{and}\,\, q_3=q_4=0,\vspace{3mm}\\
     \frac{1}{4} \left(\sqrt{36 m^2+20 m q+q^2}+6 m-q\right),\,\,\text{for}\,\,q_1=q\, \text{and}\,\, q_2=q_3=q_4=0.
\end{cases}\,
\label{phr}
\end{equation}
\end{widetext}
\begin{figure}[hbt]
       \includegraphics[scale=0.32]{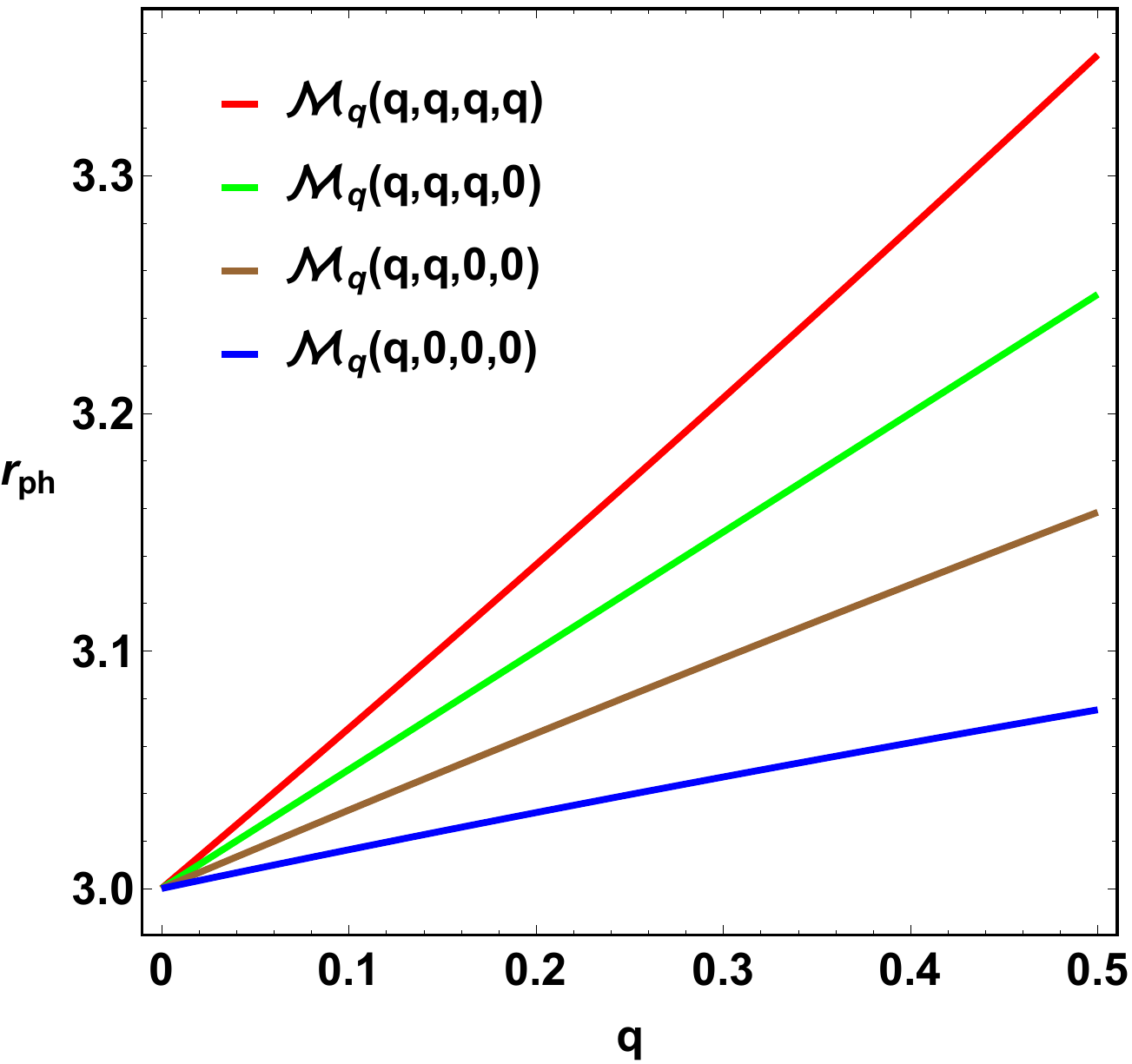}
            \caption{The photon sphere radius $r_{ph}$ versus the charge parameter $q$ for four charge configurations with $m=1$.}
            \label{Fig2}
        \end{figure}
The present step consists of defining the shape and size of the black hole, while considering the space-time $(\ref{metric})$. Toward this end, the impact parameters $\xi$ and $\eta$ are needed and related to the constants of motion $E$, $L$, and $\mathcal{C}$ by means of the following expressions:
\begin{eqnarray}
    \xi=\frac{L}{E},\quad \eta=\frac{\mathcal{C}}{E^2}.
\end{eqnarray}
In turn, the effective potential and the radial function are expressed in terms of these impact parameters by 
\begin{align}
    V_{eff}&=E^2\bigg\lbrace\frac{f(r)}{r^2\prod_{i=1}^4H_i}\left(\xi+\eta\right)-1\bigg\rbrace,\label{cons1}\\
    \mathcal{R}&=E^2\bigg\lbrace r^4\prod_{i=1}^4H_i-r^2\left(\xi+\eta\right)f(r)\bigg\rbrace.\label{cons2}
\end{align}
Afterward, injecting Eqs. $(\ref{cons1})$-$(\ref{cons2})$ into Eq. $(\ref{cons})$ leads to an equation for  $\xi$ and $\eta$ in the following compact form:
\begin{equation}
\eta+\xi^2=\frac{4r^3_{ph}\prod_{i=1}^4H_i+r_{ph}^4\left(\prod_{i=1}^4H_i\right)'}{2r_{ph}f(r_{ph})+r^2_{ph}f'(r_{ph})}.
\end{equation}
As a result, it may be correctly noted that the photon sphere $r_{ph}$ has the dimensions of  length, while the quantity $\eta+\xi^2$ has the dimensions of length square, describing a two-dimensional shadow geometry. 
 \begin{figure*}[tbh!]
    \centering
    \begin{subfigure}[b]{0.5\textwidth}
        \centering
        \includegraphics[scale=0.79]{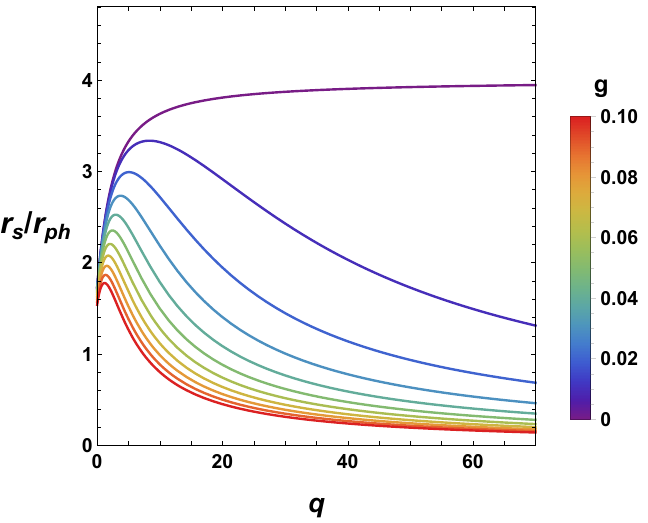}
        \caption{$q_1=q_2=q_3=q_4=q$.}
        \label{subfig:wec5}
    \end{subfigure}%
    \hfill
    \begin{subfigure}[b]{0.5\textwidth}
        \centering
        \includegraphics[scale=0.79]{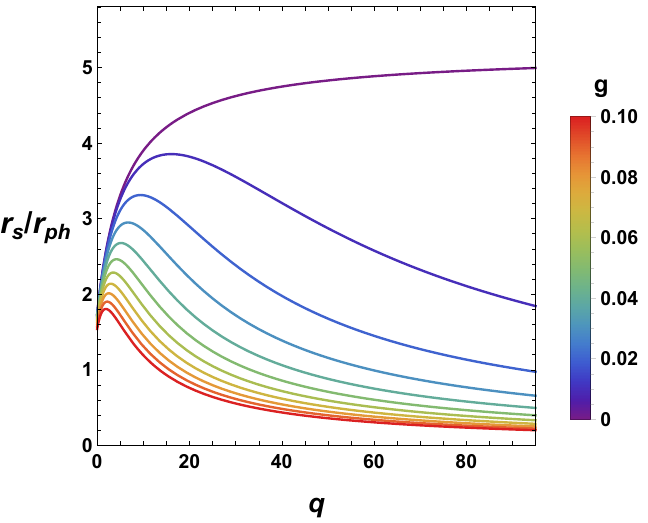}
        \caption{$q_1=q_2=q_3=q$ and $q_4=0$.}
        \label{subfig:nec5}
    \end{subfigure}%
    \\
    \begin{subfigure}[b]{0.5\textwidth}
        \centering
        \includegraphics[scale=0.79]{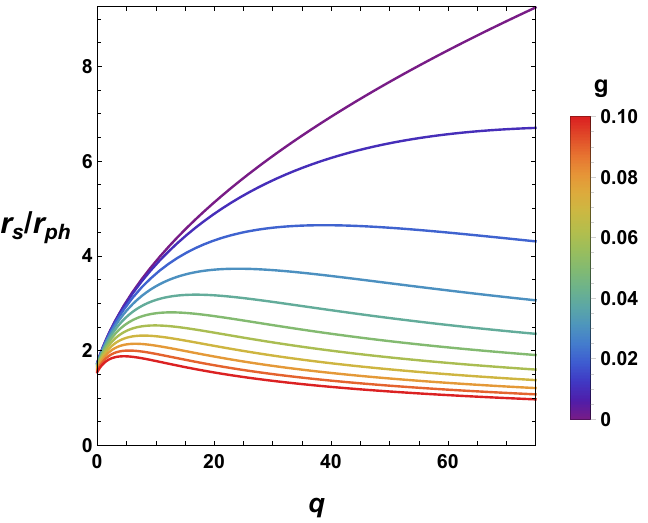}
        \caption{$q_1=q_2=q$ and $q_3=q_4=0$.}
        \label{subfig:wec6}
    \end{subfigure}%
    \hfill
    \begin{subfigure}[b]{0.5\textwidth}
        \centering
        \includegraphics[scale=0.79]{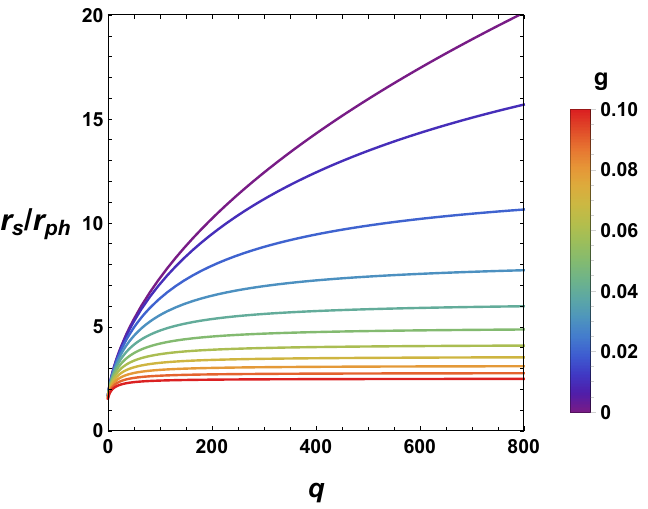}
        \caption{$q_1=q$ and $q_2=q_3=q_4=0$.}
        \label{subfig:nec6}
    \end{subfigure}%
    \caption{{\color{black}  Plot of the $r_{s}/r_{ph}$ ratio against the charge parameter $q$ for various values of the gauge coupling constant $g$.}}
    \label{Fu}
\end{figure*}

Now, we turn to reveal the visualization of the black hole shadow, i.e., the geometrical quantity on a celestial plane along the coordinates $X$ and $Y$. According to \cite{47}, the celestial coordinates are given by
\begin{align}
X&=\underset{r_O\rightarrow\infty}{lim}\left(-r_O^2\sin\theta_O\frac{d\phi}{dr}\biggr\rvert_{(r_O,\theta_O)}\right),\\
Y&=\underset{r_O\rightarrow\infty}{lim}\left(r_O^2\frac{d\theta}{dr}\biggr\rvert_{(r_O,\theta_O)}\right),
\label{E5-15}
\end{align}
where $r_O$ denotes the distance between the black hole and the observer. To be more concrete, we look at null geodesic motion in the equatorial plane {\color{black}$\theta=\pi/2$, leading} to $X=-\xi$ and $Y=\pm\sqrt{\eta}$. As a consequence, this outcome presents a two-dimensional geometry governed by the shadow radius expressed in the following way:
\begin{equation}
    r_{s}^2\equiv \eta+\xi^2=X^2+Y^2,
\end{equation}
or, in terms of the parameter space of the black hole system, one has
\begin{widetext}
\begin{equation}
    r_s^2=\frac{3 r_{ph}^2\sum_{i=1}^4q_i+2 \left(q_3 q_4+q_2 \left(q_3+q_4\right)+q_1 \left(q_2+q_3+q_4\right)\right) r_{ph}+q_1 q_2\left(q_3+q_4\right)+\left(q_1+q_2\right) q_3
   q_4+4 r_{ph}^3}{g^2 \left(3r_{ph}^2\sum_{i=1}^4q_i+2 \left(q_2 q_3+\left(q_2+q_3\right) q_4+q_1 \left(q_2+q_3+q_4\right)\right) r_{ph}+q_1 q_2\left(q_3+q_4\right)+\left(q_1+q_2\right) q_3
   q_4+4 r_{ph}^3\right)-2( m+ r_{ph})}.
   \label{shadowr}
\end{equation}
\end{widetext}
This is nothing but the shadow radius $r_s$ in celestial coordinates. For completeness, it is recalled that the shadow shape for non-rotating (static) black holes is a circle with radius  $r_s$.

       \begin{figure*}[tbh!]
    \centering
    \begin{subfigure}[b]{0.5\textwidth}
        \centering
        \includegraphics[scale=0.79]{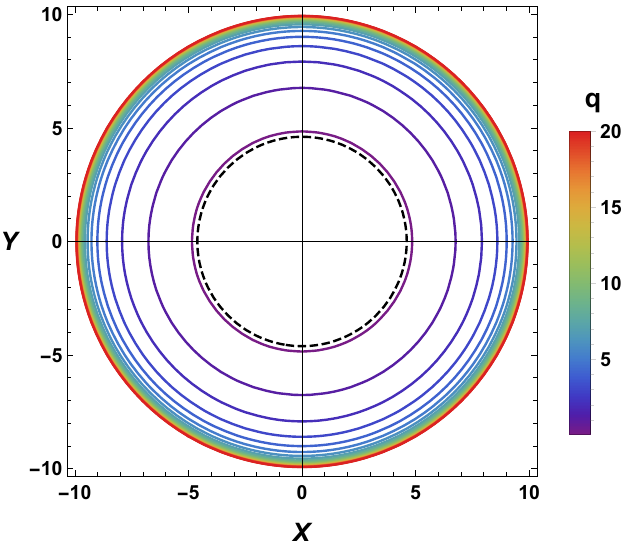}
        \caption{$q_1=q_2=q_3=q_4=q$.}
        \label{subfig:wec7}
    \end{subfigure}%
    \hfill
    \begin{subfigure}[b]{0.5\textwidth}
        \centering
        \includegraphics[scale=0.79]{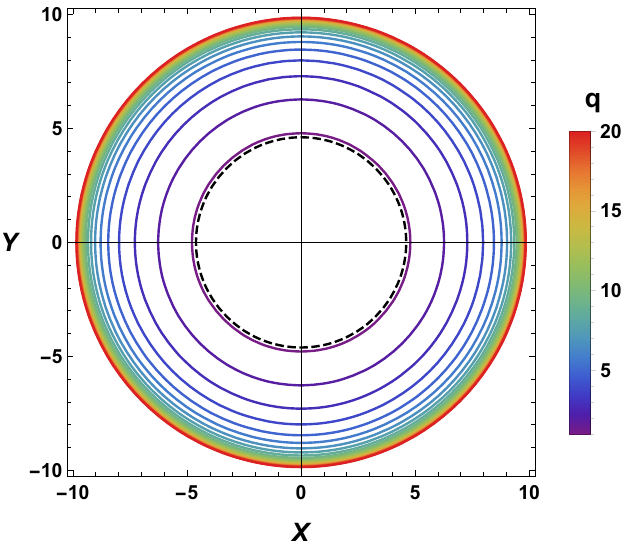}
        \caption{$q_1=q_2=q_3=q$ and $q_4=0$.}
        \label{subfig:nec7}
    \end{subfigure}%
    \\
    \begin{subfigure}[b]{0.5\textwidth}
        \centering
        \includegraphics[scale=0.79]{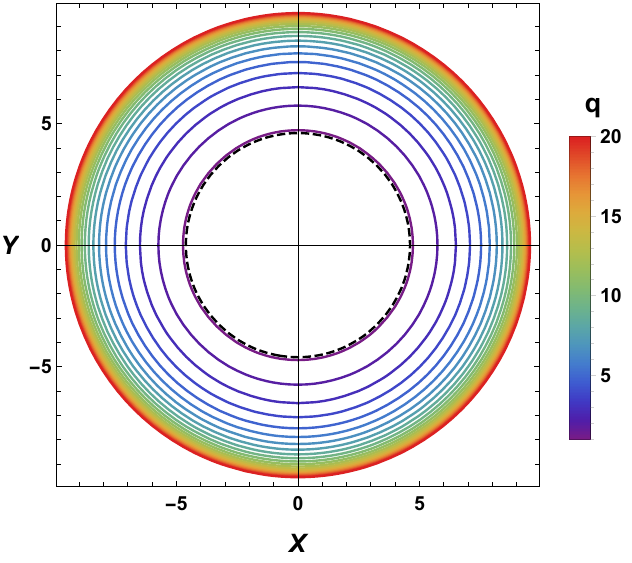}
        \caption{$q_1=q_2=q$ and $q_3=q_4=0$.}
        \label{subfig:wec8}
    \end{subfigure}%
    \hfill
    \begin{subfigure}[b]{0.5\textwidth}
        \centering
        \includegraphics[scale=0.79]{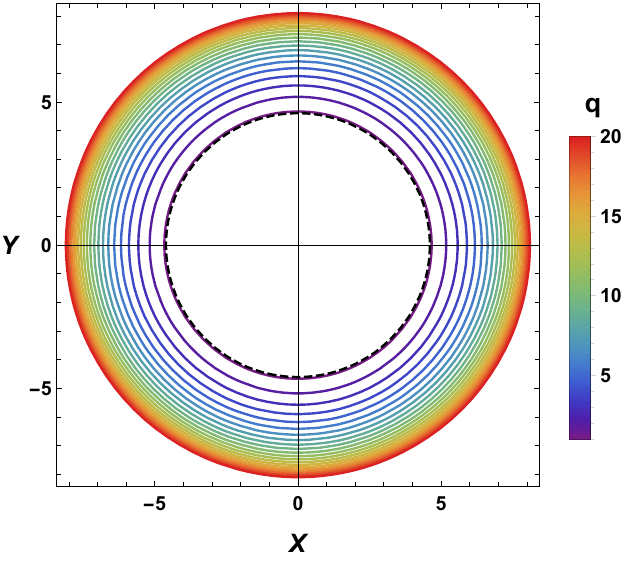}
        \caption{$q_1=q$ and $q_2=q_3=q_4=0$.}
        \label{subfig:nec8}
    \end{subfigure}%
            \caption{The geometrical shape of the shadow of the $\mathbb{S}\mathbb{T}\mathbb{U}$  black hole in the celestial plane regarding four charge configurations with $m = 1$ and $g=0.1$.}
            \label{Fig3}
        \end{figure*}
To better understand the photon sphere radius depends on the parameter space, we plot in Fig. (\ref{Fig2}) its behavior as a function of $q$ and for the four different charge configurations indicated above. It is observed that the photon sphere exhibits a positive proportionality with the variation in electric charge $(q)$ and the existence of a common origin point at $q = 0$.

{\color{black} The plots in Fig.~\ref{Fu} display the behavior of the ratio between the shadow radius $r_{s}$ and the photon sphere radius $r_{ph}$ against the charge parameter $q$, for various values of the gauge coupling constant $g$. It is noted that the cases of four, three, two equal charges, and a single charge indicate a maximal value of the $r_{s}/r_{ph}$ ratio in correspondence of the minimal value of the gauge coupling constant (i.e. $g=0$ - no coupling), and vice-versa. 
As a function of the entire value of the gauge coupling $g$, the plots also show that the lower bound of the ratio $r_{s}/r_{ph}$ remains nearly constant, in contrast to the upper bound, which tends to increase just for cases 3 and 4. On the other hand, revealing the lower and upper bounds of the charge parameter concerning the $r_{s}/r_{ph}$ ratio with a given coupling constant is illustrated for four cases of the parameter space. A closer look shows that the $r_{s}/r_{ph}$ ratio for all cases in parameter space has a unique value at $q=0$, whatever the value of the gauge coupling constant $g$. This is a natural intuition that has already been demonstrated in the results mentioned above. The study of the lower and upper bounds of the charge parameter $q$ is discussed by analyzing the behavior of the ratio $r_{s}/r_{ph}$ with the considered gauge coupling interval $0.01\leq g \leq0.1$ except for the ungauged case, which has no maximum. Consequently, all the cases involve two phases, namely increasing and decreasing the ratio $r_{s}/r_{ph}$ except the last one with a single charge, which does not present any maximum value of the ratio $r_{s}/r_{ph}$. The maximum of the ratio $r_{s}/r_{ph}$ according to the studied cases and taken into account the interval $0.01\leq g \leq0.1$ may be examined as follows 
\begin{itemize}
    \item case 1: $q_1=q_2=q_3=q_4=q$.
         \begin{eqnarray}
              1.18443&\leq& q_{max}\leq 8.34627,\\1.7752&\leq&\left(\frac{r_{s}}{r_{ph}}\right)^{max} \leq3.3339.
         \end{eqnarray}
     \item case 2:  $q_1=q_2=q_3=q$ and $q_4=0$.
         \begin{eqnarray}
              1.89869&\leq& q_{max}\leq 16.0961,\\
1.80189&\leq&\left(\frac{r_{s}}{r_{ph}}\right)^{max} \leq3.851.
         \end{eqnarray}
     \item case 3: $q_1=q_2=q$ and $q_3=q_4=0$.  
    \begin{eqnarray}
         4.5838&\leq& q_{max}\leq 84.4517,\\
1.8757&\leq&\left(\frac{r_{s}}{r_{ph}}\right)^{max} \leq6.7137.
    \end{eqnarray}
\end{itemize}
Moreover, a further analysis shows the maximal value of the charge parameter in which all $g$-curves coincide at a unique point. So, the computations indicate that: 
\begin{itemize}
\item case 1:  $q^{maxi}=897.611 $, 
\item case 2: $q^{maxi}=1793.45 $, 
\item  case 3: $q^{maxi}=807300 $ ,
\end{itemize}
whereas the case of a single charge entails a comparatively larger maximal value on $q$. Briefly, the $r_{s}/r_{ph}$ ratio presents significant constraints on the minimal and maximal values of the two parameters, namely the gauge coupling $g$ and the electric charge $q$.}

\begin{figure}[H]
       \includegraphics[scale=0.5]{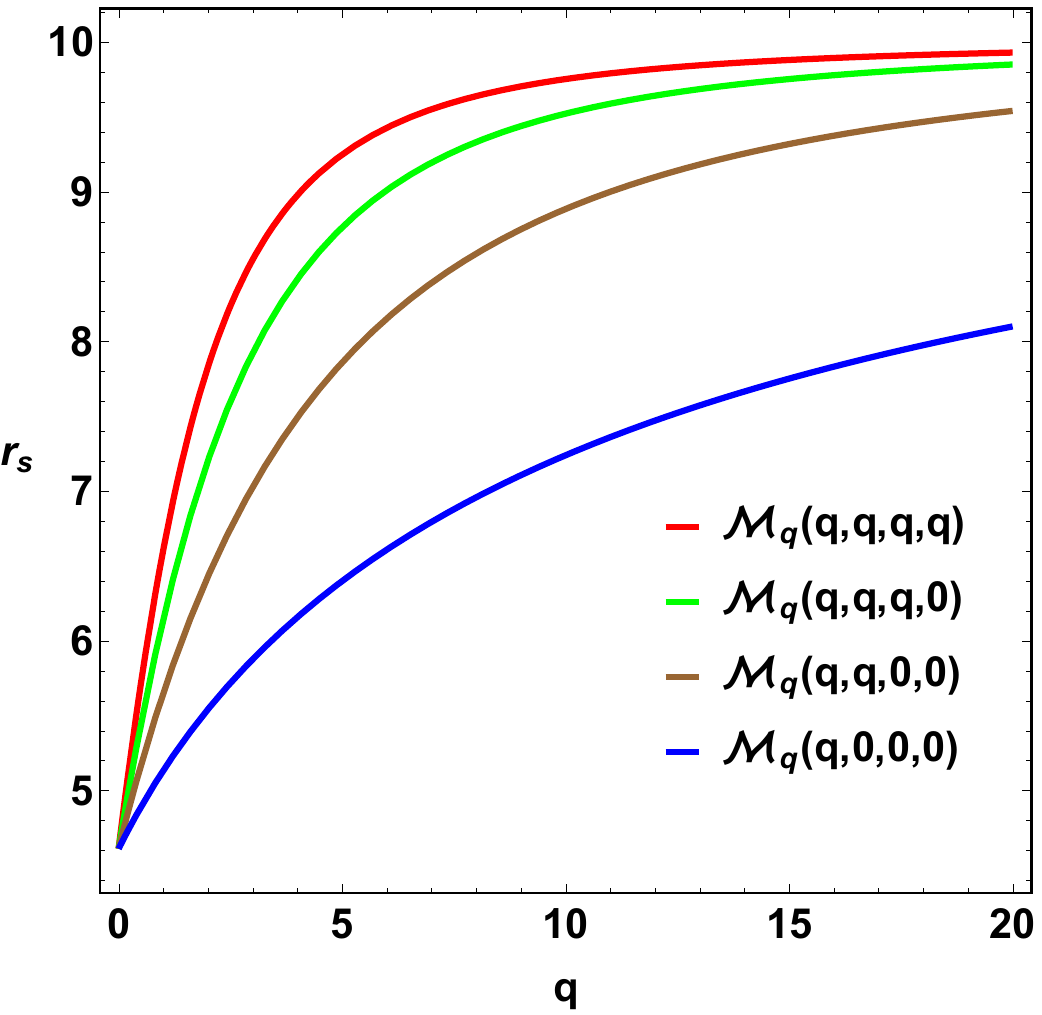}
            \caption{Variation of the shadow radius $r_s$ as a function of the charge parameter $q$ regarding four cases of charge configurations with $m=1$ and $g=0.1$}
            \label{Fig4}
        \end{figure}

Now, the plots in Fig.~$\ref{Fig3}$ provide appropriate observations to study the optical shadow aspect.  First and foremost, the investigation is conceptualized in terms of the four cases of charge configurations. Accordingly, the four figures point out two relevant aspects: on the one hand, the effect of the charge parameter on the shadow behavior in the four cases, especially the size of the shadow. On the other hand,  the effect of the parameter space $\mathcal{M}_q\left(q_1,q_2,q_3,q_4\right)$ on the size of the shadow. Indeed, as the charge value increases, the shadow size increases. In addition, all cases share the same shadow radius for the uncharged case $(q=0)$, which is the small concentric circle (the black dashed circle). From the point of view of the parameter space, it is clear from Fig.~$\ref{Fig4}$ that the size of the shadow decreases once the parameter space of the four electric charges is reduced. Besides, as the number of charges increases, the number of photon spheres increases.

\subsection{Energy emission rate}
\label{S4fig}
\begin{figure*}[tbh!]
    \centering
    \begin{subfigure}[b]{0.5\textwidth}
        \centering
        \includegraphics[scale=0.79]{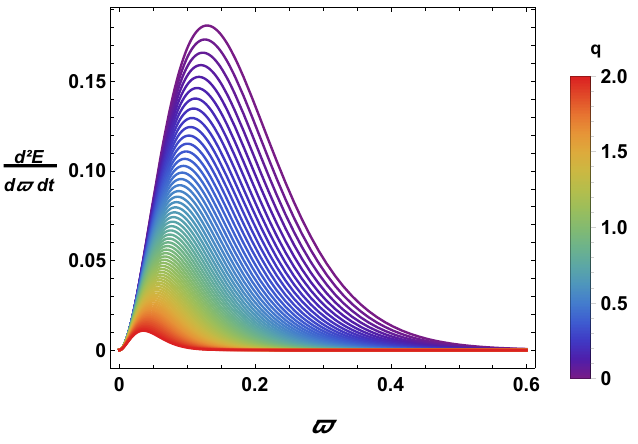}
        \caption{$q_1=q_2=q_3=q_4=q$.}
        \label{subfig:wec9}
    \end{subfigure}%
    \hfill
    \begin{subfigure}[b]{0.5\textwidth}
        \centering
        \includegraphics[scale=0.79]{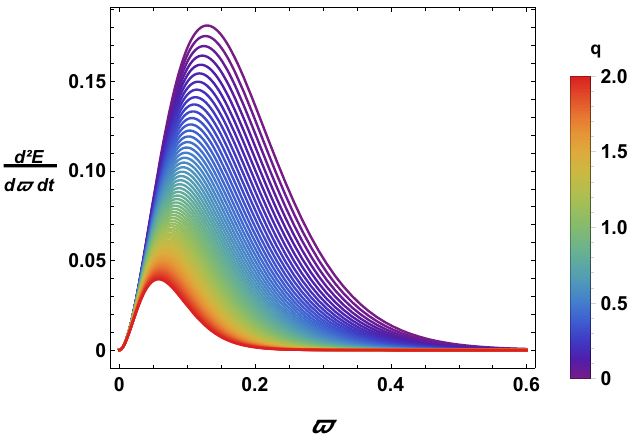}
        \caption{$q_1=q_2=q_3=q$ and $q_4=0$.}
        \label{subfig:nec9}
    \end{subfigure}%
    \\
    \begin{subfigure}[b]{0.5\textwidth}
        \centering
        \includegraphics[scale=0.79]{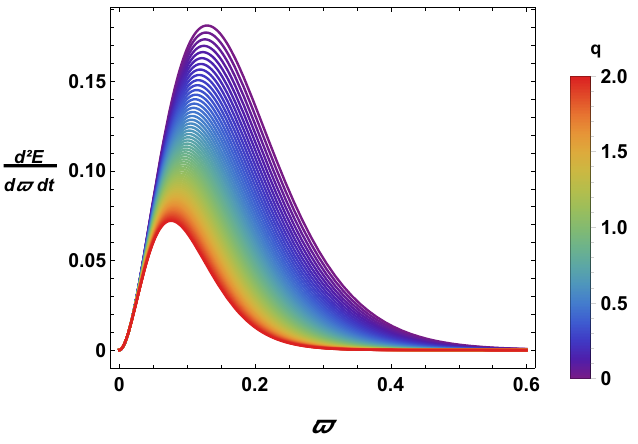}
        \caption{$q_1=q_2=q$ and $q_3=q_4=0$.}
        \label{subfig:wec10}
    \end{subfigure}%
    \hfill
    \begin{subfigure}[b]{0.5\textwidth}
        \centering
        \includegraphics[scale=0.79]{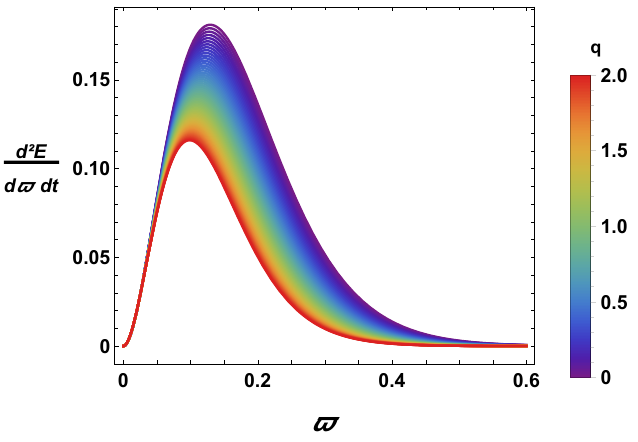}
        \caption{$q_1=q$ and $q_2=q_3=q_4=0$.}
        \label{subfig:nec10}
    \end{subfigure}%
            \caption{The energy emission rate as a function of $\varpi$ for the $\mathbb{S}\mathbb{T}\mathbb{U}$ black hole for four cases of charge configurations with $g= 0.1$}
            \label{Fig5}
        \end{figure*}

The main objective of this section is to analyze the behavior of the energy emission rate. In addition, due to the process leading to the formation and annihilation of certain pairs of particles near the horizons of the black hole, particles of positive energy can escape from the black hole via tunnels located in the region where Hawking radiation occurs. This process, known as Hawking radiation, results in the evaporation of the black hole over time. The absorption cross-section has been shown to approach the shadow of the black hole for a distant observer \cite{48}. At very high energies, the absorption cross-section behaves like an oscillator with a limit constant value of ($\sigma_{lim}=\pi r_s^2$). The corresponding energy emission rate can be expressed as follows \cite{48,49,50,51}:
\begin{equation}
    \frac{d^2E(\varpi)}{dtd\varpi}=\frac{2\pi^3\varpi^3 r_s^2}{e^{\frac{\varpi}{T}}-1},
    \label{E5-16}
\end{equation}
in which $\varpi$ is the emission frequency, and $T$ is the Hawking
temperature given by
\begin{equation}
    T=\frac{2 g r_h^3\sum_{i=1}^4q_i +\mathcal{V} \,r_h^2+3 g r_h^4-g \prod_{i=1}^4q_i}{4
   \pi  r_h^3 \sqrt{\prod_{i=1}^4\left(1+\frac{q_i}{r_h}\right)}},\label{tu}
\end{equation}
where
\begin{equation}
    \mathcal{V}=g \,q_3 q_4+g q_2 \left(q_3+q_4\right)+g q_1 \left(q_2+q_3+q_4\right)+1.
\end{equation}
To inspect graphically the energy emission rate behavior, Fig.~$\ref{Fig5}$ shows the variation of the energy emission rate as a function of $\varpi$ for four cases of charge configurations. It is worth noting that for all four cases, the charge parameter influences the variation of the energy emission rate in such a way that once the charge increases, the energy emission rate decreases. As a consequence of the inconsistency between the variation of the charge parameter and the energy emission rate, the candidate scenario points to a black hole evaporation process that is slower and has an extended lifetime.

{\color{black}
        \subsection{Sparsity}
This section looks at the phenomenon of sparsity for the $\mathbb{S}\mathbb{T}\mathbb{U}$ black holes. Roughly speaking, this phenomenon is a specific aspect of black hole radiation. In practical terms, sparsity is defined as the average time between the emission of successive quanta on time scales set by the energies of the emitted quanta. A key feature of Hawking radiation is that it emerges as extremely sparse throughout the evaporation process, compared with black body radiation, enabling us to differentiate between the two phenomena.

Sparsity can be quantified by introducing the parameter~\cite{Sp1,Sp2}
\begin{equation}
\label{defSpars}
    \eta =\frac{C}{\Tilde{g} }\left(\frac{\lambda_t^2}{\mathcal{A}_{eff}}\right),
\end{equation}
where $C$ is a dimensionless constant, $\Tilde{g}$ denotes the spin degeneracy factor of the emitted quanta, $\lambda_t = 2\pi/T$ is nothing more than the thermal wavelength, and $\mathcal{A}_{eff}= 27\mathcal{A} /4$ represents the effective area of the BH. For Schwarzschild BHs, the emission of massless bosons is explicitly $\lambda_t = 2\pi/T_{Sch} = 8\pi^2r_h$ leading to the following result:
\begin{equation}
   \eta_{Sch} = \frac{64 \pi ^3}{27}\approx  73.49 \gg 1.
\end{equation}
Notably, for a black body case, the situation  $\eta\ll1$ is perfectly provided.

To proceed with revealing the physics profile of the sparsity aspect, we consider four cases of charge configurations. This process is carried out by considering the corresponding temperature (\ref{tu}) and taking into account the area of the horizon, $\mathcal{A}=\pi\sqrt{\prod_{i=1}^4\left(r_h+q_i\right)}$. Detailed expressions of the sparsity in terms of the full parameter space of the four abelian gauge fields are found  in App. \ref{A2}.

It is useful to think about the sparsity limits that give rise to defined reduced sparsity, essentially in terms of $g$ and the charge $q$. Notice that the setting $g =q = 0$ can lead to the definition of the sparsity of the Schwarzschild BH. Correspondingly, in the configurations of four equal charges (case 1) and two equal charges (case 3), both at the $(g\rightarrow0)$ limit, the sparsity can be specifically outlined as follows 
\begin{align}
     \eta(g\rightarrow0)&=\eta_{Sch} \left(1+\frac{2q}{r_h}\right),\qquad \text{case 1},\\
      \eta(g\rightarrow0)&=\eta_{Sch} \left(1+\frac{q}{r_h}\right)\qquad \text{case 3},
\end{align}
which tend to coincide with the exact sparsity of the Schwarzschild BH whenever the model becomes uncharged $\left( q = 0\right)$, i.e., $\eta=\eta_{Sch}$. However, the unpair case of the charge configuration provides an exact limit without giving rise to the charge parameter $q$, given as
\begin{align}
     \eta(g\rightarrow0)&=\eta_{Sch}.
\end{align}

In order to obtain an accurate interpretation of the sparsity behavior in all four charge configuration options, Fig.~$\ref{Spa}$ displays the relevant outline. It is observed for the initial phase that sparsity has a higher function for small horizon radii. This behavior is consistent with all charge configurations. Furthermore, for the next phase, the scenario is very close to the divergence of the sparsity at a specified horizon radius. This behavior is remarkably similar in each charge configuration case. In other words, the available behavior of the sparsity presents two physical scenarios involving Hawking radiation. Broadly stated, within the first phase, at a small horizon along with a larger sparsity value, Hawking radiation is actually not a continuous emission of particles but perfectly closely mimics sparse radiation, i.e., most particles are emitted randomly and discretely, with gaps between them. However, in the second phase, at a large horizon radius, the spacing function collapses asymptotically. This infers that Hawking radiation acts like blackbody radiation, in which the thermal wavelength is much shorter than the size of the emitting body.

A closer inspection, regarding a comparative study between the four cases of the charge configuration, shows the following results: all the cases share the same behavior of the sparsity function in relation to the charge variation. So, as the charge decreases, the transition from the sparse radiation profile to the black body radiation is faster. Further, the horizontal line displayed in all cases of the charge configurations exactly represents the Schwarzschild case over the selection $\left(g =0, q = 0\right)$.
\begin{figure*}[tbh!]
    \centering
    \begin{subfigure}[b]{0.5\textwidth}
        \centering
        \includegraphics[scale=0.75]{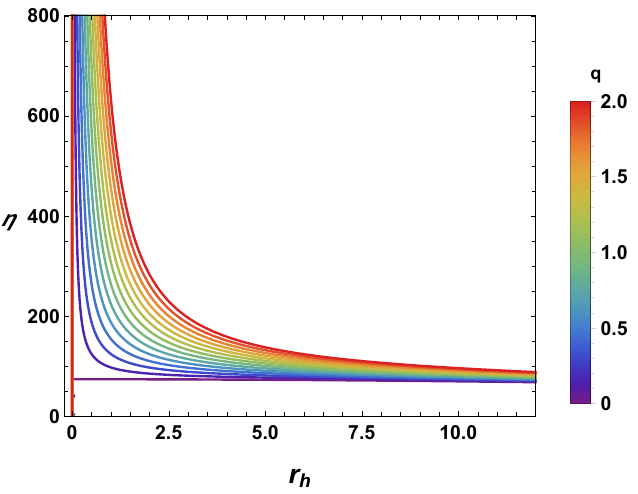}
        \caption{$q_1=q_2=q_3=q_4=q$.}
        \label{subfig:wec11}
    \end{subfigure}%
    \hfill
    \begin{subfigure}[b]{0.5\textwidth}
        \centering
        \includegraphics[scale=0.75]{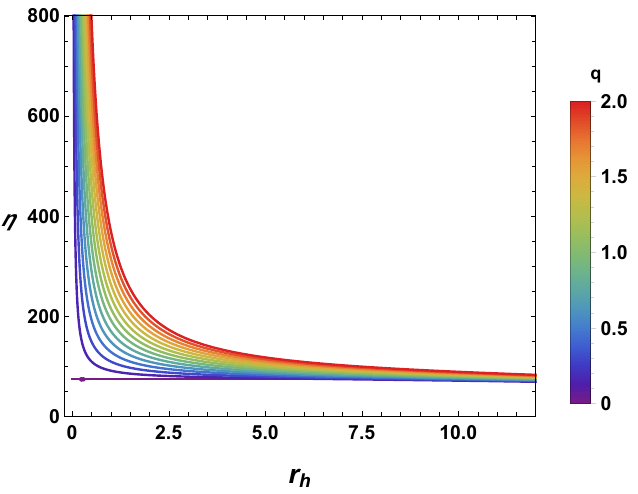}
        \caption{$q_1=q_2=q_3=q$ and $q_4=0$.}
        \label{subfig:nec11}
    \end{subfigure}%
    \\
    \begin{subfigure}[b]{0.5\textwidth}
        \centering
        \includegraphics[scale=0.75]{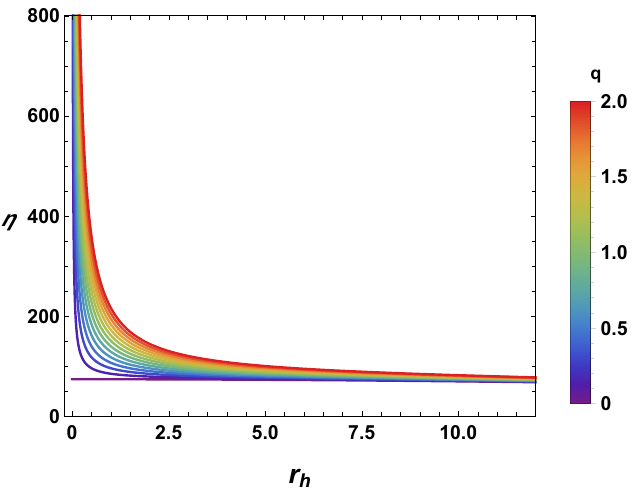}
        \caption{$q_1=q_2=q$ and  $q_3=q_4=0$.}
        \label{subfig:wec12}
    \end{subfigure}%
    \hfill
    \begin{subfigure}[b]{0.5\textwidth}
        \centering
        \includegraphics[scale=0.75]{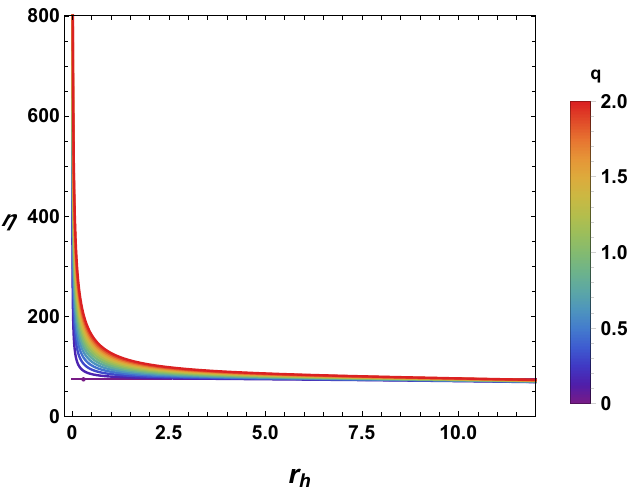}
        \caption{$q_1=q$ and $q_2=q_3=q_4=0$.}
        \label{subfig:nec12}
    \end{subfigure}
           \caption{{\color{black}Plot of $\eta$ as a function of $r_h$ for four cases of charge configurations with $g =0.01.$}} \label{Spa}
        \end{figure*}
        }
        
\section{Constraints on shadow from M87$^\star$ and Sgr A$^\star$}
\label{S4}

By taking into account the spatial separation \(D\) between the supermassive black hole (SMBH) and the galactic center, the classical diameter of the shadow can be determined through the application of the customary arclength equation:

\begin{equation} \label{earc}
d_\text{sh} = \frac{D \theta_\text{sh}}{M}.
\end{equation}

The computed measurements for the shadow diameter of M87$^\star$ is as follows: it exhibits a diameter of \(d^\text{M87*}_\text{sh} = (11 \pm 1.5)M\) \cite{EventHorizonTelescope:2019dse, EventHorizonTelescope:2022wkp}. In this instance, we have utilized the uncertainties documented in Refs. \cite{EventHorizonTelescope:2021dqv, Vagnozzi:2022moj} to determine restrictions on the values of black hole charge $q$. It is worth highlighting that these uncertainties are more stringent in comparison to those resulting from the use of Equation \eqref{earc}. Interested readers are encouraged to consult the aforementioned references to gain insight into the methodology employed to derive the $1\sigma$ and $2\sigma$ confidence levels. In this work, we shall calculate the $1\sigma$ bounds on the black hole charge using the results of M87$^\star$ {\color{black} and Sgr A$^\star$}.

\begin{figure*}[tbh!]
    \centering
    \begin{subfigure}[b]{0.5\textwidth}
        \centering
        \includegraphics[scale=0.6]{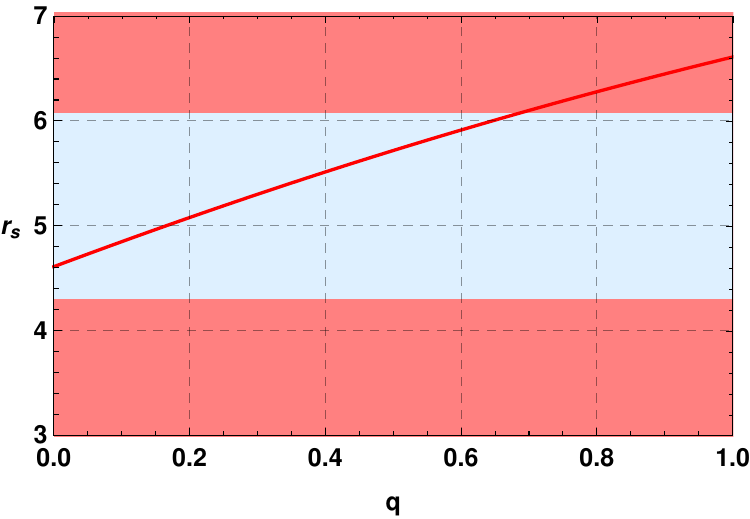}
        \caption{$q_1=q_2=q_3=q_4=q$.}
        \label{subfig:wec13}
    \end{subfigure}%
    \hfill
    \begin{subfigure}[b]{0.5\textwidth}
        \centering
        \includegraphics[scale=0.6]{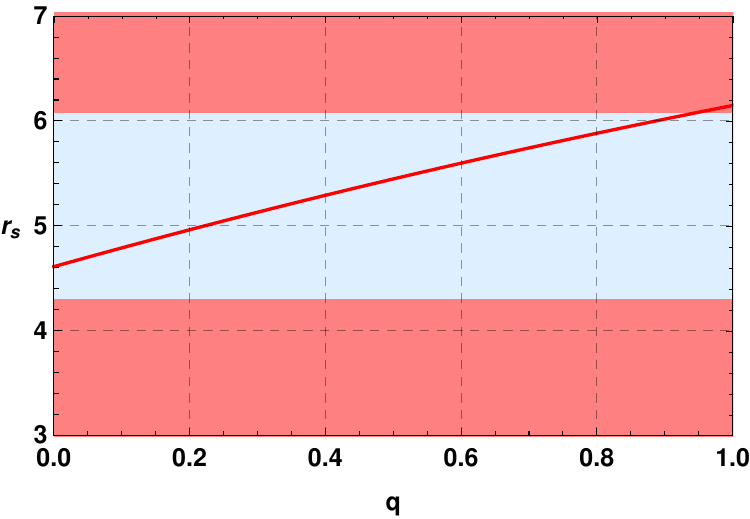}
        \caption{$q_1=q_2=q_3=q$ and $q_4=0$.}
        \label{subfig:nec13}
    \end{subfigure}%
    \\
    \begin{subfigure}[b]{0.5\textwidth}
        \centering
        \includegraphics[scale=0.6]{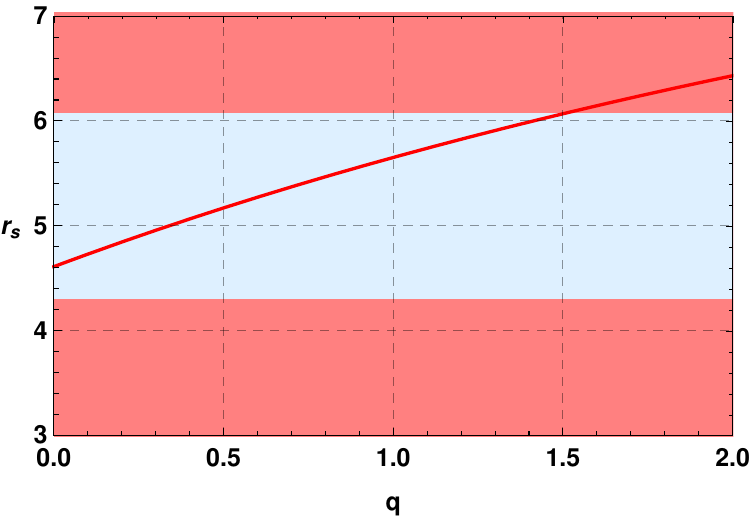}
        \caption{$q_1=q_2=q$ and  $q_3=q_4=0$.}
        \label{subfig:wec14}
    \end{subfigure}%
    \hfill
    \begin{subfigure}[b]{0.5\textwidth}
        \centering
        \includegraphics[scale=0.6]{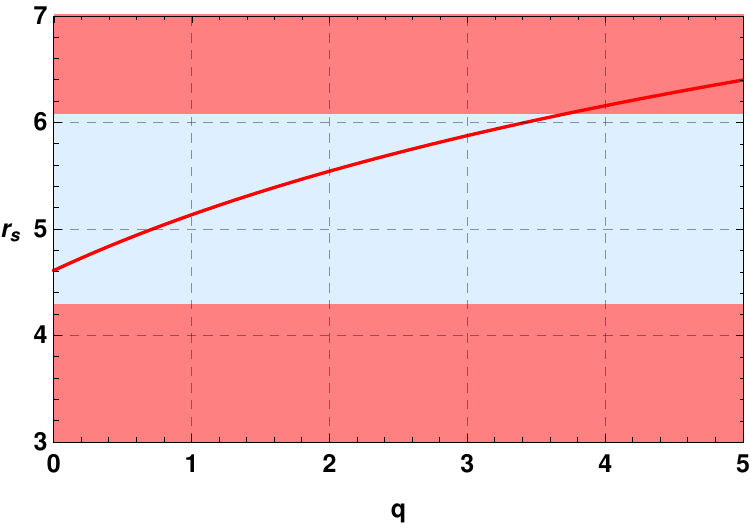}
        \caption{$q_1=q$ and $q_2=q_3=q_4=0$.}
        \label{subfig:nec14}
    \end{subfigure}%
    \caption{ $1\sigma$ constraints on the black hole shadow using M87$^\star$.}
    \label{constraint01}
\end{figure*}

\begin{table*}[!ht]
    \centering
    \begin{tabular}{clllll}
\hline
\hline
{$q$} &    Case 1 & Case 2 & Case 3 & Case 4 \\
\hline
$1\sigma$(upper/lower) from M87$^\star$    &       $0.688898$ / none &       $0.94758$ / none & $1.51472$ / none & $3.70475$ / none \\
\hline
\end{tabular}
\caption{Values of $q$ based on the constraints imposed by the EHT data on the shadow radius as shown in Fig. \ref{constraint01}.}
    \label{constrainttab}
\end{table*}

\begin{figure*}[tbh!]
    \centering
    \begin{subfigure}[b]{0.5\textwidth}
        \centering
        \includegraphics[scale=0.6]{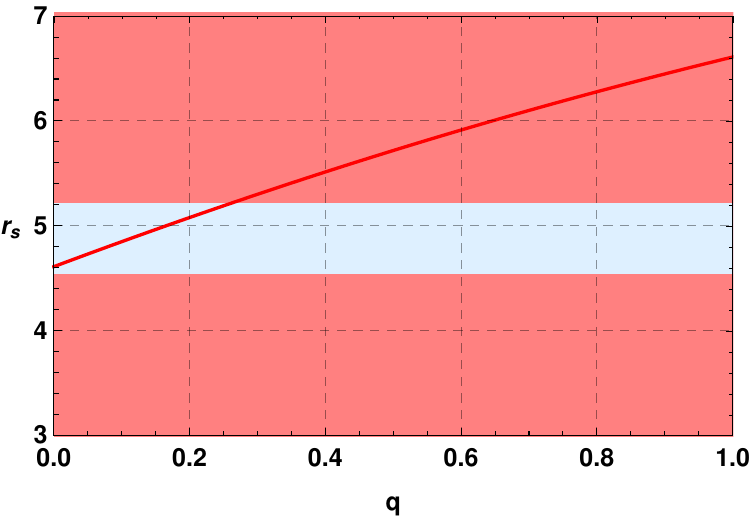}
        \caption{$q_1=q_2=q_3=q_4=q$.}
        \label{subfig:wec15}
    \end{subfigure}%
    \hfill
    \begin{subfigure}[b]{0.5\textwidth}
        \centering
        \includegraphics[scale=0.6]{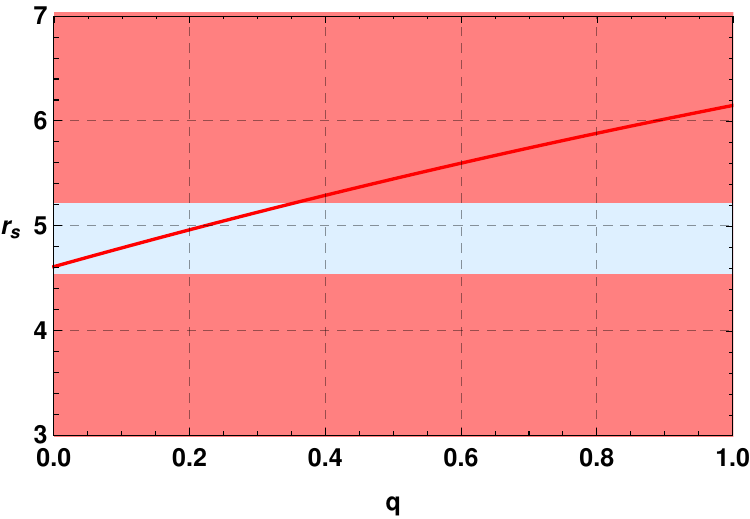}
        \caption{$q_1=q_2=q_3=q$ and $q_4=0$.}
        \label{subfig:nec15}
    \end{subfigure}%
    \\
    \begin{subfigure}[b]{0.5\textwidth}
        \centering
        \includegraphics[scale=0.6]{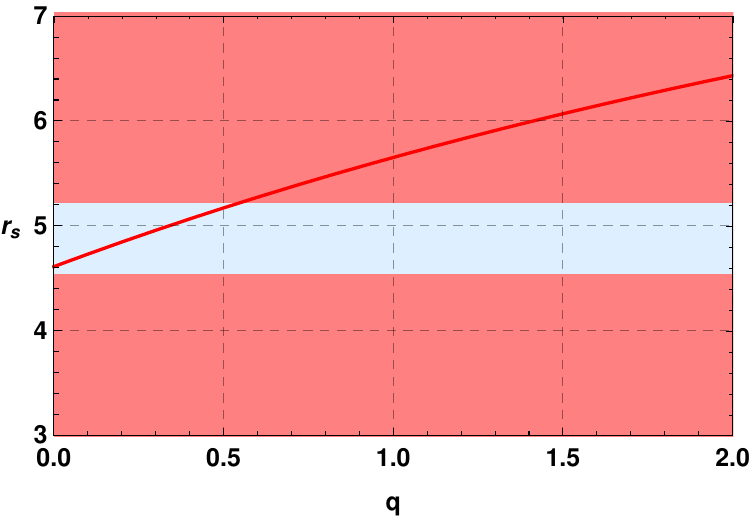}
        \caption{$q_1=q_2=q$ and  $q_3=q_4=0$.}
        \label{subfig:wec16}
    \end{subfigure}%
    \hfill
    \begin{subfigure}[b]{0.5\textwidth}
        \centering
        \includegraphics[scale=0.6]{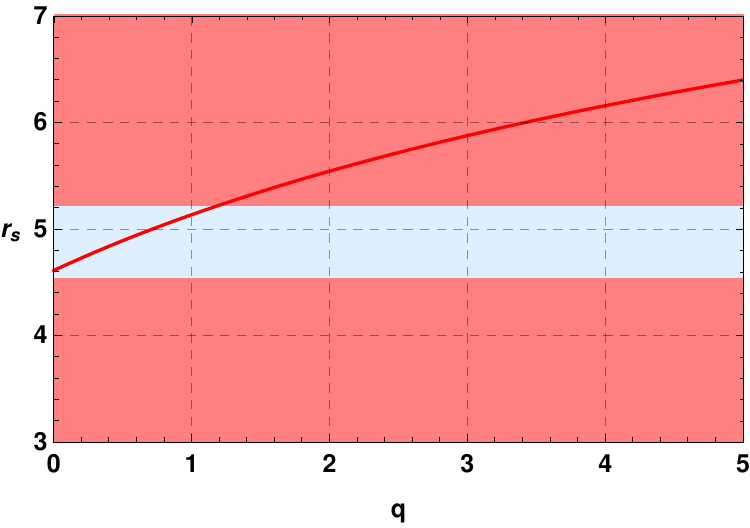}
        \caption{$q_1=q$ and $q_2=q_3=q_4=0$.}
        \label{subfig:nec16}
    \end{subfigure}%
    \caption{ \color{black} $1\sigma$ constraints on the black hole shadow using Sgr A$^\star$.}
    \label{constraint02}
\end{figure*}

\begin{table*}[!ht]
    \centering
    \begin{tabular}{clllll}
\hline
\hline
{$q$} &    Case 1 & Case 2 & Case 3 & Case 4 \\
\hline
$1\sigma$(upper/lower) from Sgr A$^\star$    &       $0.263118$ / none &       $0.355577$ / none & $0.54803$ / none & $1.19093$ / none \\
\hline
\end{tabular}
\caption{ \color{black} Values of $q$ based on the constraints imposed by the Sgr A$^\star$ observational data on the shadow radius as shown in Fig. \ref{constraint02}.}
    \label{constrainttab2}
\end{table*}

Our analysis presents the potential limitations imposed on the black hole charge $q$ within the $1\sigma$ confidence interval of the shadow radius of M87$^\star$. The outcomes are visually represented in Fig. \ref{constraint01}, and the corresponding constraints on the charge $q$ are tabulated in Table \ref{constrainttab}. The investigation reveals that the lower $1\sigma$ boundary of the black hole's shadow radius $r_s$ does not impose any restrictions on the charge $q$. However, the upper $1\sigma$ boundary of $r_s$ restricts the maximum allowable value of the black hole charge $q$ for all four cases. Specifically, in case 1, the imposed bound is relatively smaller, whereas in case 4 the constraint on $q$ is more extensive.

{\color{black} Constraints on $q$ from observational data of Sgr A$^\star$ is shown graphically in Fig. \ref{constraint02} and the constraint values of $q$ are shown in Table \ref{constrainttab2}. One may note that Sgr A$^\star$ data can constrain the charge parameter more stringently compared to the previous study.}

The constraint on the black hole shadow can be further tested with the observational results of quasinormal modes of black holes. One may note that the black hole shadow is related to the real part of the quasinormal modes of a black hole. Hence, in the near future, with significant observational data of quasinormal modes from space-based gravitational wave detectors like LISA \cite{LISA:2017pwj}, it could be possible to obtain another set of constraints on the model parameters and the consistency between the theories can be tested.

\section{Phase structure of the $\mathbb{S}\mathbb{T}\mathbb{U}$ black hole using shadow analysis}
\label{S5}
To better investigate thermodynamic phase structure from the perspective of shadow analysis, we follow the analysis of \cite{522}, which provides descriptive scenarios on the subject. For this purpose, we restrict the study to two cases of charge configurations, namely four equal charges (case 1) and three equal charges (case 2).

\begin{figure*}[hbt]
      	\centering{
      	\includegraphics[scale=0.8]{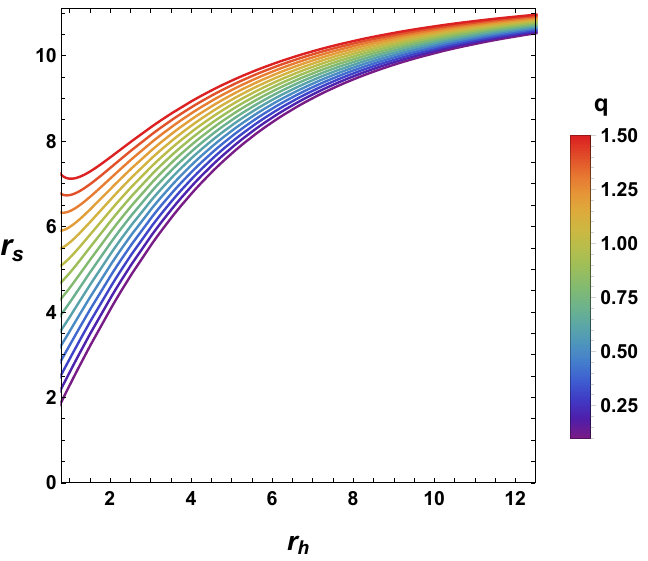}
       \includegraphics[scale=0.8]{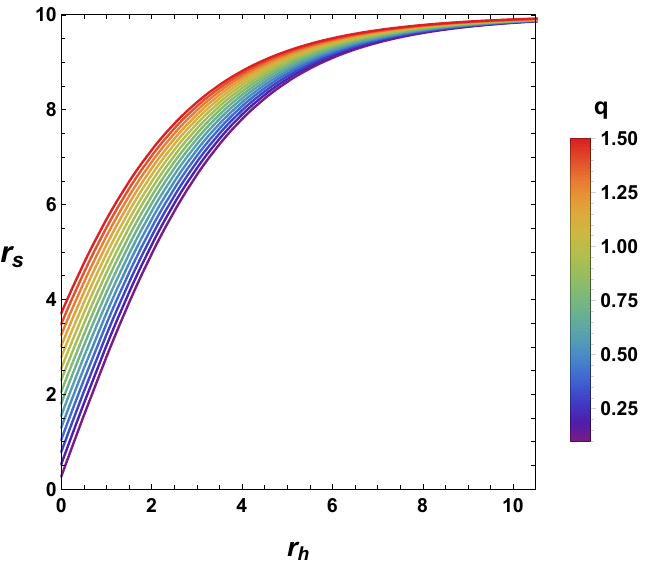}
     
       }
      	\caption{The variation of shadow radius $r_s$ against the event horizon radius $r_h$, here we have set $g = 0.1$. Left graph: $\mathcal{M}_q\left(q,q,q,q\right)$, Right graph: $\mathcal{M}_q\left(q,q,q,0\right)$.}
      	\label{Fig6}
      \end{figure*}

To study the phase structure features of the $\mathbb{S}\mathbb{T}\mathbb{U}$ black hole, it is required to define the bare mass parameter $\left( m\right)$ in terms of the radius of the event horizon $r_h$. Equating $f(r=r_h)$ defined in Eq. $(\ref{Eq6})$ to zero, one obtains
\begin{equation}
    m=\frac{r_h}{2}\bigg(1+\frac{g^2}{r_h^2}\prod_{i=1}^4\left(q_i+r_h\right)\bigg).
    \label{mass}
\end{equation}
To discover the link between the shadow radius $(r_s)$ and the horizon radius $(r_h)$, we first consider Eq. $(\ref{shadowr})$ and then take the two solutions of Eq. $(\ref{phr})$ given by
\begin{widetext}
    \begin{equation}
r_{ph}= \begin{cases}
     \frac{1}{2} \left(\sqrt{9 m^2+2 m q+q^2}+3 m+q\right),\,\,\,\,\,\,\text{for}\,\,q_1=q_2=q_3=q_4=q, \vspace{3mm}\\
    \frac{1}{2} (6 m+q),\,\,\qquad\qquad\qquad\qquad\qquad\,\,\,\,\text{for}\,\,q_1=q_2=q_3=q\, \text{and}\,\, q_4=0.\vspace{3mm}
\end{cases}\,
\label{II}
\end{equation}
\end{widetext}
By plugging Eq. $(\ref{mass})$ into Eq. $(\ref{II})$, one obtains an expression unifying the photon sphere radius $(r_s)$ and the horizon radius $(r_h)$, which is injected into Eq.$(\ref{shadowr})$ to obtain the desired results, i.e.,
\begin{widetext}
    \begin{equation}
    r_s^2=\begin{cases}
        \frac{r_h(q+r_h)^3}{r_h^2-g^2(q-3r_h)(q+r_h)^3},\,\,\,\,\,\,\,\,\,\qquad\qquad\text{for}\,\,q_1=q_2=q_3=q_4=q,
        \vspace{5mm}\\
        \frac{27(q+r_h)^2(1+g^2(q+r_h)^2)^2}{(1+3g^2(q+r_h)^2)^2(4+3g^2(q+r_h)^2)},\,\,\quad\,\,\,\,\text{for}\,\,q_1=q_2=q_3=q\, \text{and}\,\, q_4=0,
    \end{cases}\,
    \label{RSRH}
\end{equation}
\end{widetext}
which are nothing more than the expression of the $r_s-r_h$ relation for the cases of four and three equal charges, respectively. The comparison between the two cases in relation to the horizon radius and the shadow radius is depicted in Fig.~$\ref{Fig6}$. Inspection shows proportionality between the horizon radius and the shadow radius, implying the existence of a positive correlation. Accordingly, the shadow radius can play a relevant role in reflecting the thermodynamic phase transition of the $\mathbb{S}\mathbb{T}\mathbb{U}$ black holes. The findings are perfectly similar to several studies of thermodynamics from the point of view of shadow analysis \cite{52,23}.

The present step consists of revealing the critical behavior on the $T$-$r_h$ diagram and the heat capacity phase transition in terms of the shadow radius. The isotherm curve $P$-$r_h$ is obtained by considering both Eqs. $(\ref{Eq10})$-$(\ref{P})$. Thus, the related state equation is given by
\begin{equation}
    T=\frac{r_h^4+\frac{4}{3}\pi P\left(-\prod_{i=1}^3q_i+r_h^4\sum_{i=1}^3q_i+2r_h^6\right)}{2\pi r_h^5\sqrt{\prod_{i=1}^3\left(1+\frac{q_i}{r_h^2}\right)}}.
\end{equation}
In turn, the inflection (i.e. critical) point is defined by imposing the following conditions:
\begin{align}
    \frac{\partial T}{\partial r_h}\bigg\rvert_{P=P_c}=0,\,\,
    \frac{\partial^2 T}{\partial r_h^2}\bigg\rvert_{P=P_c}=0.
\end{align}
Accordingly, the values of the critical radius $rc$, temperature $T_c$ and pressure $P_c$ for the charge configurations in question are explicitly obtained as follows:
\begin{itemize}
    \item case 1: $q_1=q_2=q_3=q_4=q$,
\begin{center}
\hspace{9mm}$r_c=2.1139\,q$ $\quad$ $T_c=\frac{0.0245143}{q}$ $\quad$ $P_c=\frac{0.00136599}{q^2}$.
\end{center}
    \item case 2: $q_1=q_2=q_3=q$ and $q_4=0$,
\begin{center}
\hspace{9mm}$r_c=1.15139\,q$ $\quad$ $T_c=\frac{0.0377334}{q}$ $\quad$ $P_c=\frac{0.00339089}{q^2}$.
\end{center}
\end{itemize}
{\color{black}It is observed that critical pressure and temperature are decreasing functions of the charge parameter $q$, whereas the critical horizon radius is an increasing function of it. Furthermore, the critical set of pressure and temperature for case 2 is greater than that of case 1, while the critical horizon radius for case 2 is smaller than that of case 1.

On the other hand, cases 3 and 4 do not exhibit any critical points, implying that no transition can occur in these two cases.} 

\begin{figure*}[hbt]
      	\centering{
      	\includegraphics[scale=0.51]{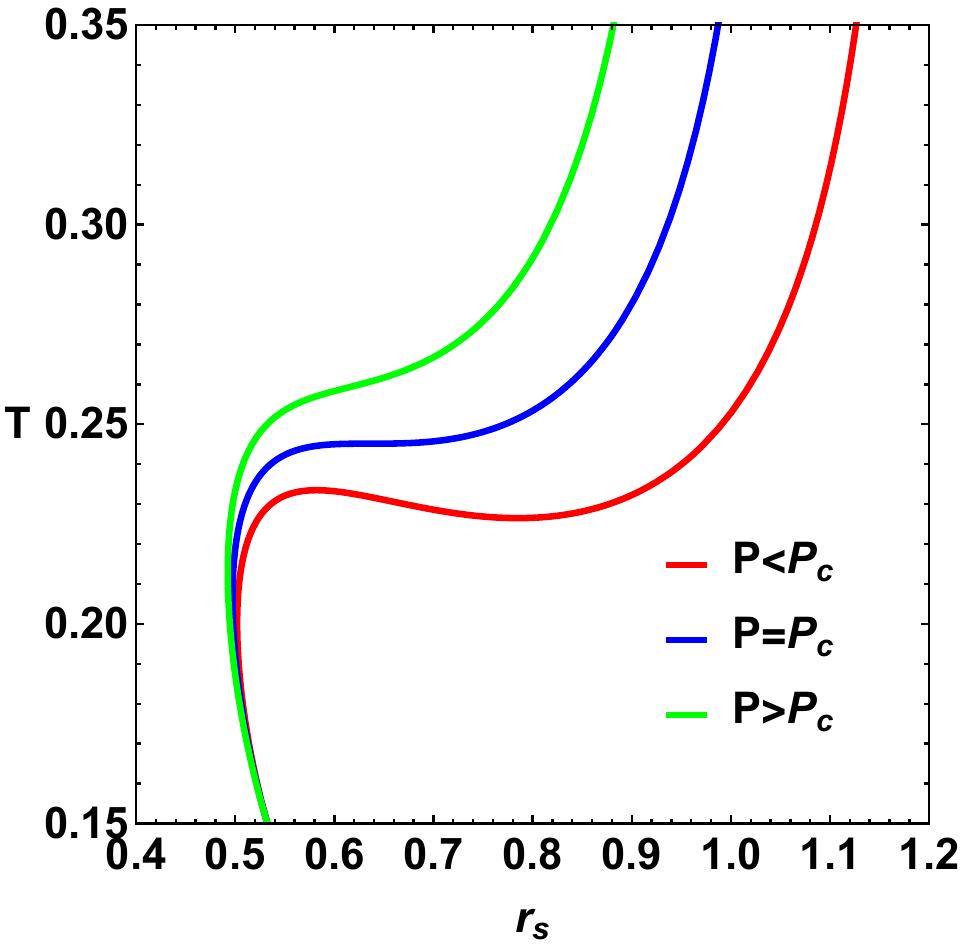}
       \includegraphics[scale=0.50]{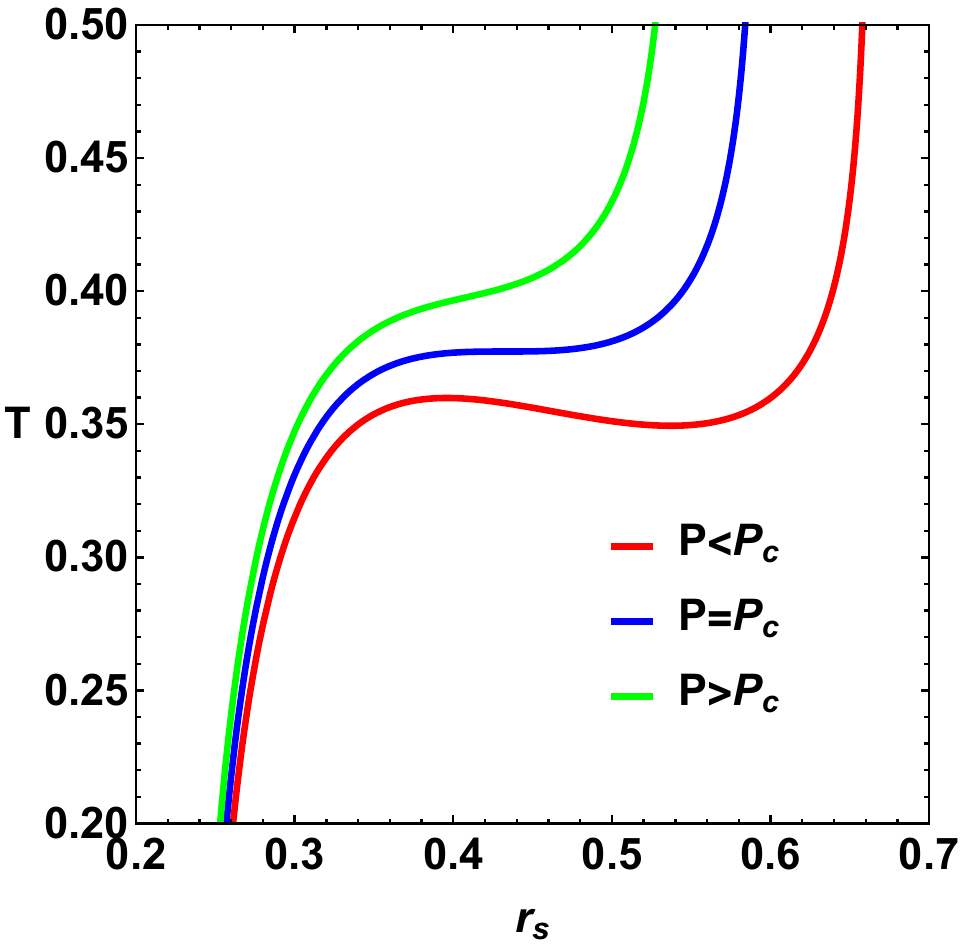}
    
       }
      	\caption{Hawking temperature with respect to the shadow radius $r_s$, Left graph: temperature $T$ in terms of $r_s$ for four equal charges; Right graph: temperature $T$ in terms of $r_s$ for three equal charges.}
      	\label{Fig7}
      \end{figure*}

{\color{black}To highlight the key properties of the $T$-$r_s$ behavior, Fig.~$\ref{Fig7}$ shows the isobaric curve on the diagram $T-r_s$, which is classified according to multiple pressure values. Indeed, the appropriate case $P<P_c$ implies two extreme points along the sides of three branches, namely the small black hole branch, the intermediate black hole branch, and the large black hole branch. The small and large black hole branches are characterized by a positive slope, which means that the heat capacity is positive and the system is thermally stable (see also the discussion below Eq.~\eqref{Cp}). On the other hand, the intermediate black hole branch stands on a negative slope, which entails that the black hole is thermally unstable due to the negativity of the heat capacity. By contrast, the case $P>P_c$ indicates the absence of any extremal point, which proves that the system exhibits a unique stable branch in this situation.}
      
To better explore the local stability from the point of view of the shadow analysis, we consider the expression of the heat capacity as a function of the shadow radius, which is given by 
\begin{equation}
\label{Cp}
    C_{P, q}=T\bigg(\frac{\partial S}{\partial r_h}\frac{\partial r_h}{\partial T}\bigg).
\end{equation}
To point out whether the system is locally stable or unstable, the sign of the heat capacity provides information on the thermal stability and phase transition of the black hole system, as explained above. In particular, a positive (negative) sign claims a thermally stable (unstable) system. Indeed, $\frac{\partial S}{\partial r_h} > 0$, which means that the sign of $C_{P,q}$ depends on $\frac{\partial T}{\partial r_h}$. 
Clearly, the term $\frac{\partial T}{\partial r_h}$ can be further expanded as follows:
\begin{equation}
    \frac{\partial T}{\partial r_h}=\frac{\partial T}{\partial r_s}\frac{\partial r_s}{\partial r_h}.
\end{equation}
\begin{figure*}[hbt]
      	\centering{
      	\includegraphics[scale=0.52]{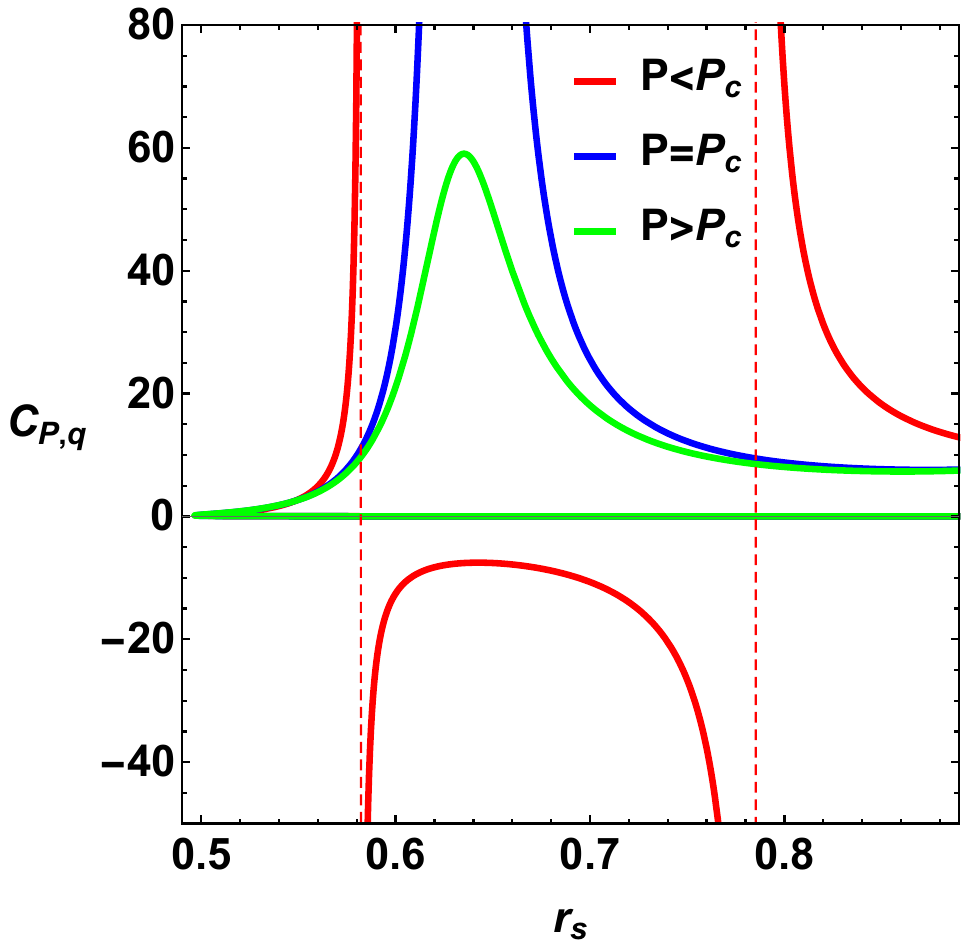}
       \includegraphics[scale=0.52]{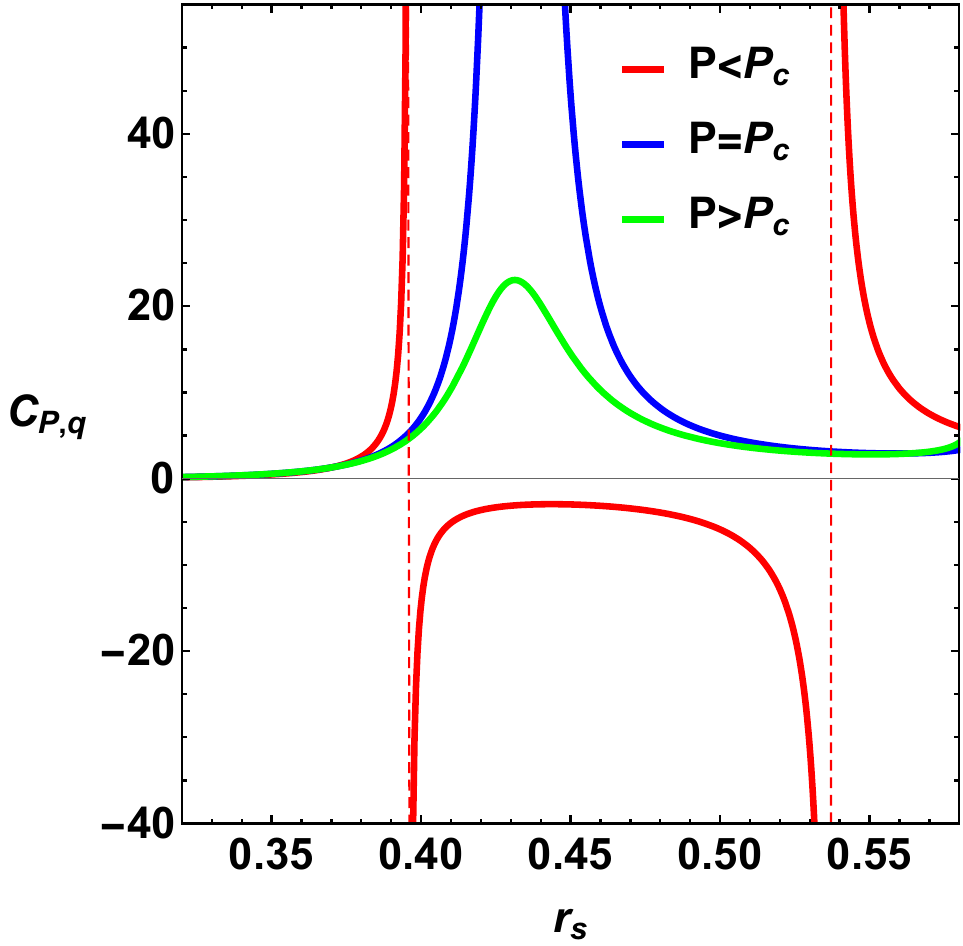}
       }
      	\caption{Heat capacity with respect to the shadow radius $r_s$. Left graph: $C_{P,q}$ in terms of $r_s$ for four equal charges; Right graph: $C_{P,q}$ in terms of $r_s$ for three equal charges. }
      	\label{Fig8}
      \end{figure*}
Since $\frac{\partial r_s}{\partial r_h}>0$, which ensures a positive correlation between the shadow radius and the event horizon radius, the sign of $C_{P,q}$ is always inspected by $\frac{\partial T}{\partial r_s}$. The variation in heat capacity $C_{P,q}$ versus shadow radius $r_s$ is shown in Fig.~$\ref{Fig8}$ for different values of pressure. For $P<P_c$, there are two critical points where the heat capacity diverges and changes signs, giving rise to a second-order phase transition of the black hole. As the pressure increases to the critical value $P_c$, the number of critical points is reduced to one, while above $P_c$ no critical point is reached and no phase transition occurs.

 {\color{black}In summary and in terms of benchmark, since $T_c (\text{case 2}) >T_c (\text{case 1})$, the inflection point in case 2 takes a higher value than in case 1 (Fig. \eqref{Fig7}). This reflects the effect of the reduced parameter space of the charge on the phase transition points. Remarkably, a similar outcome has been found in~\cite{Gogoi:2021syo} in the case of a five-dimensional R-charged black hole. It can also be seen that the two physically divergent points are shifted to the left relative to case 1 as a result of reducing the parameter space to three equal charges. In addition, the physical region lying in the small and large black hole phase transition in the space of $r_s$ tends to enlarge as the configuration moves from case 1 to case 2.}

\begin{figure*}[tbh!]
      	\centering{
  {\includegraphics[width=.31\textwidth]{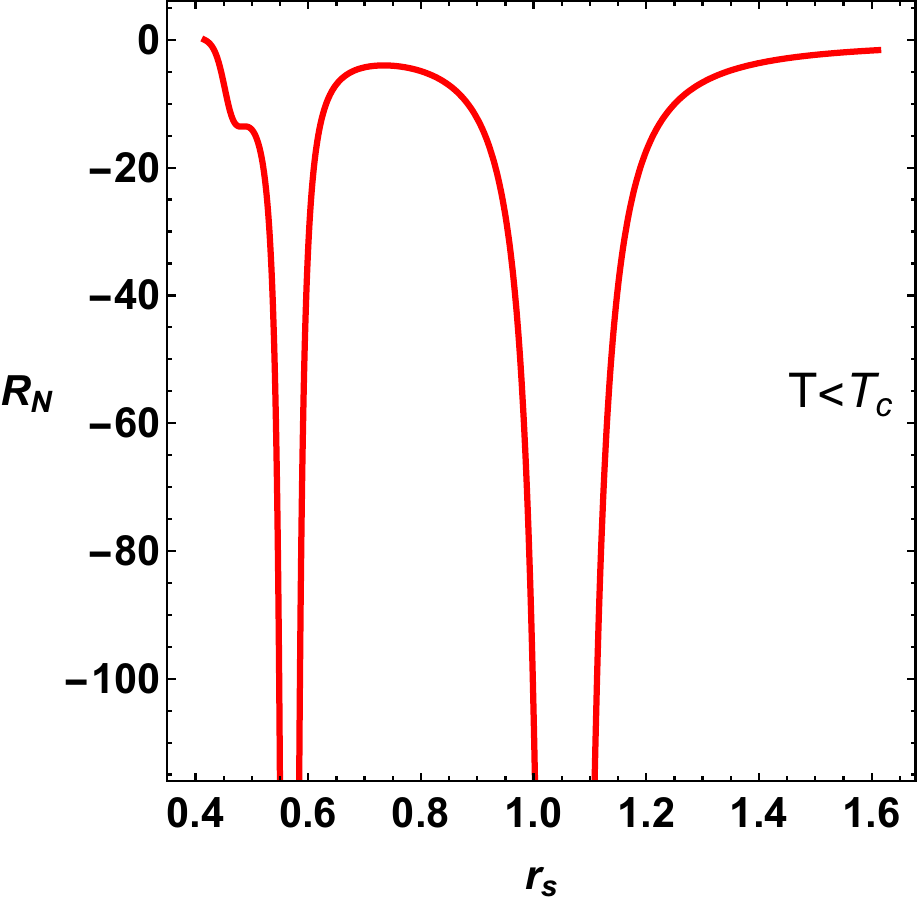}}
  {\includegraphics[width=.31\textwidth]{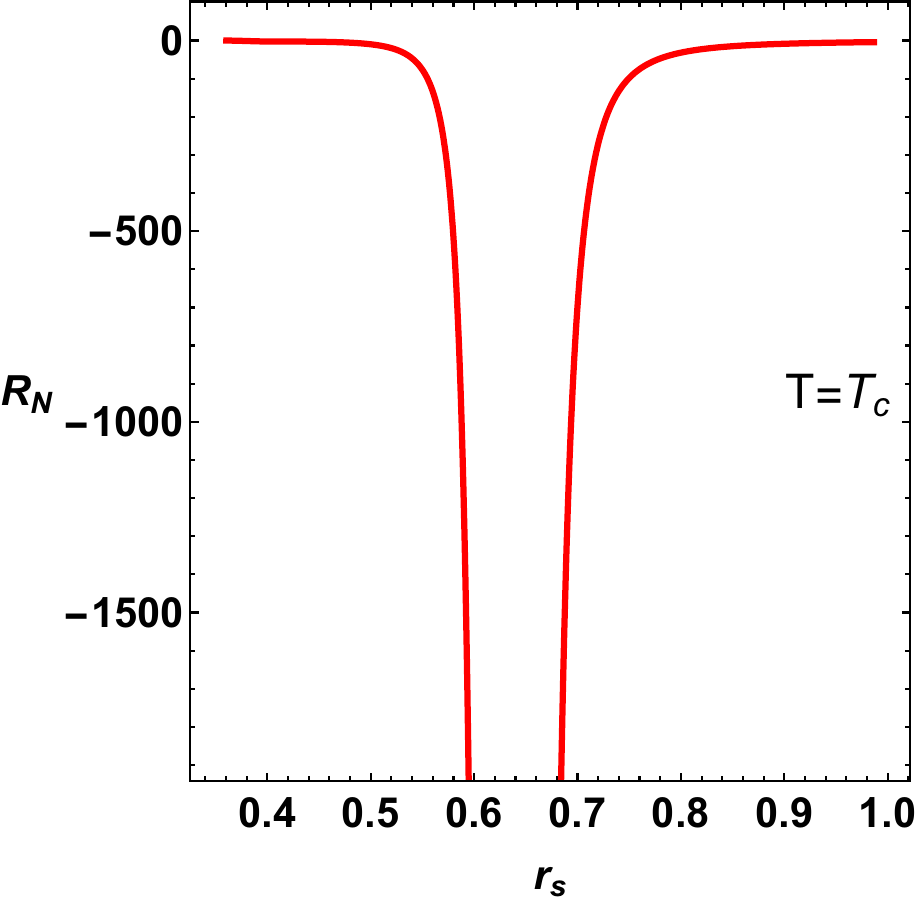}}
  {\includegraphics[width=.31\textwidth]{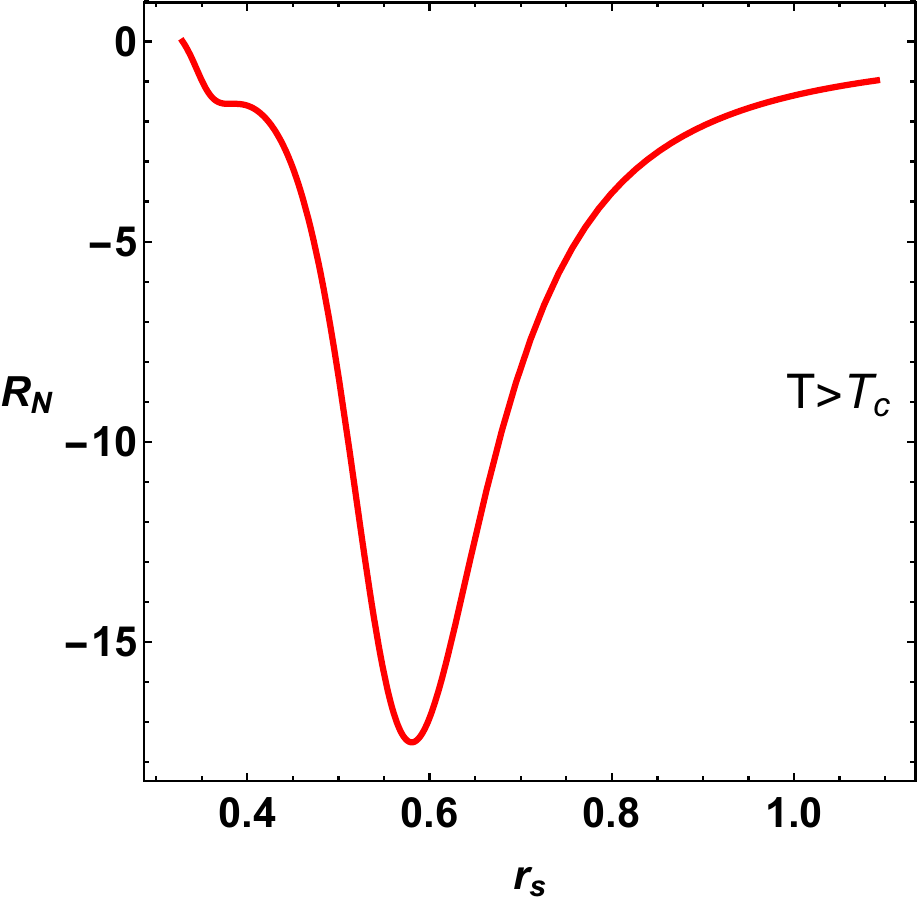}}\par
  {\includegraphics[width=.31\textwidth]{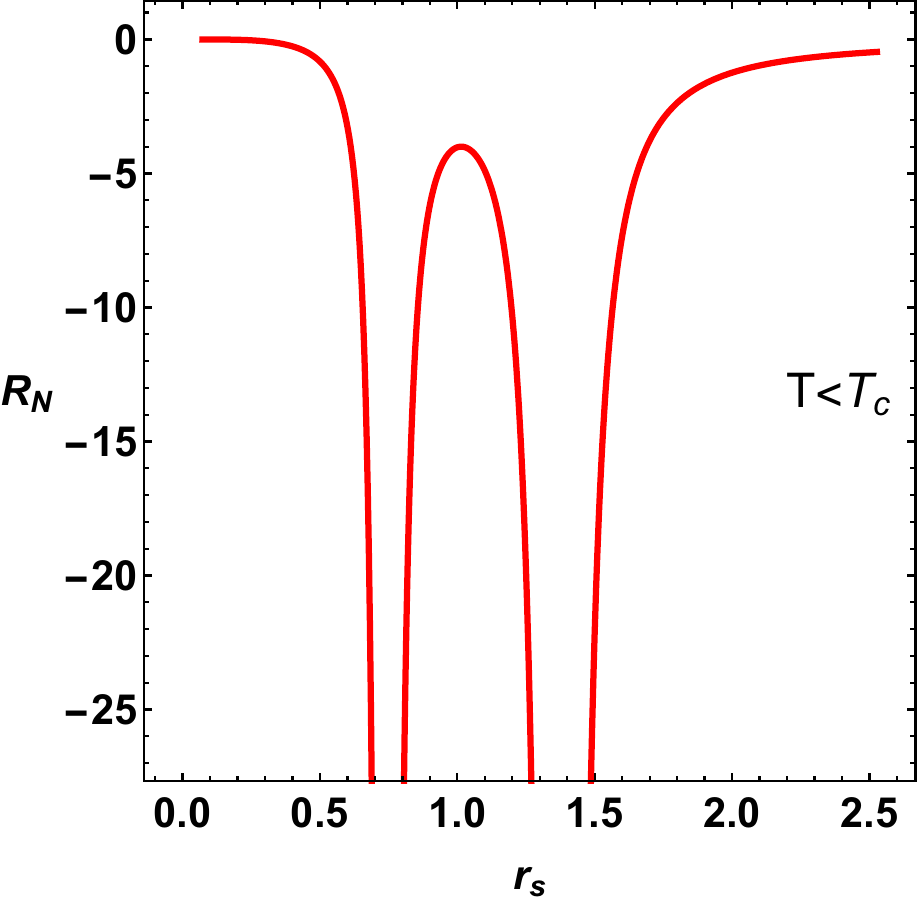}}
  {\includegraphics[width=.31\textwidth]{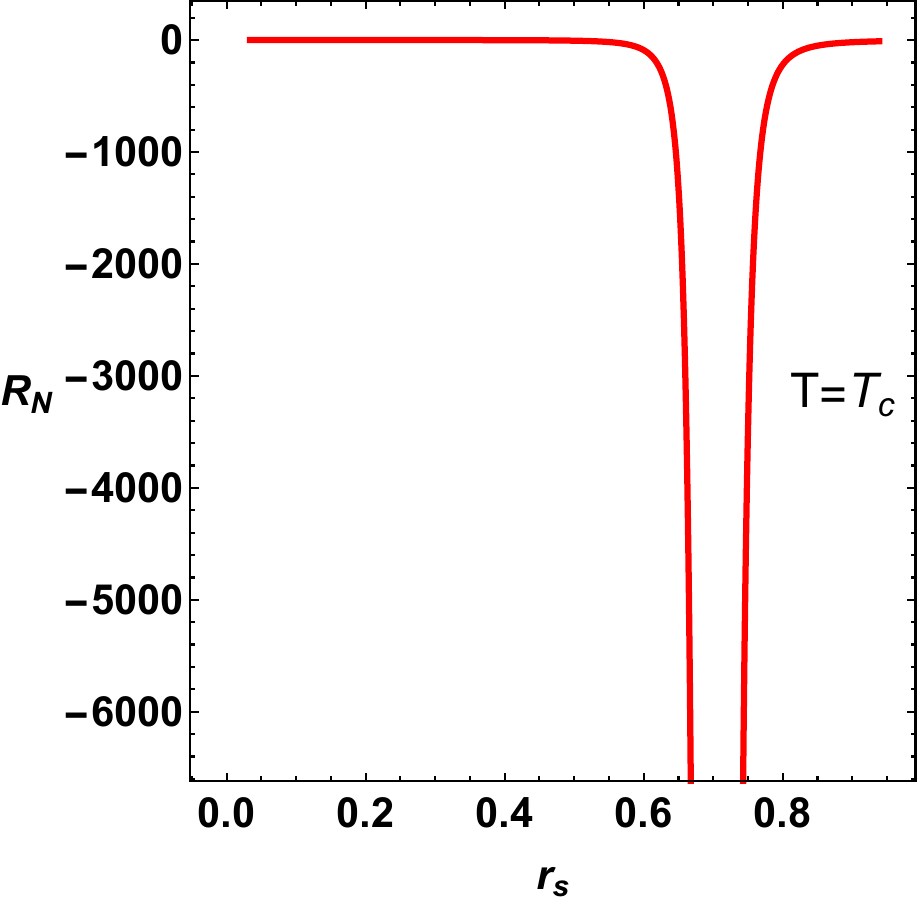}}
  {\includegraphics[width=.31\textwidth]{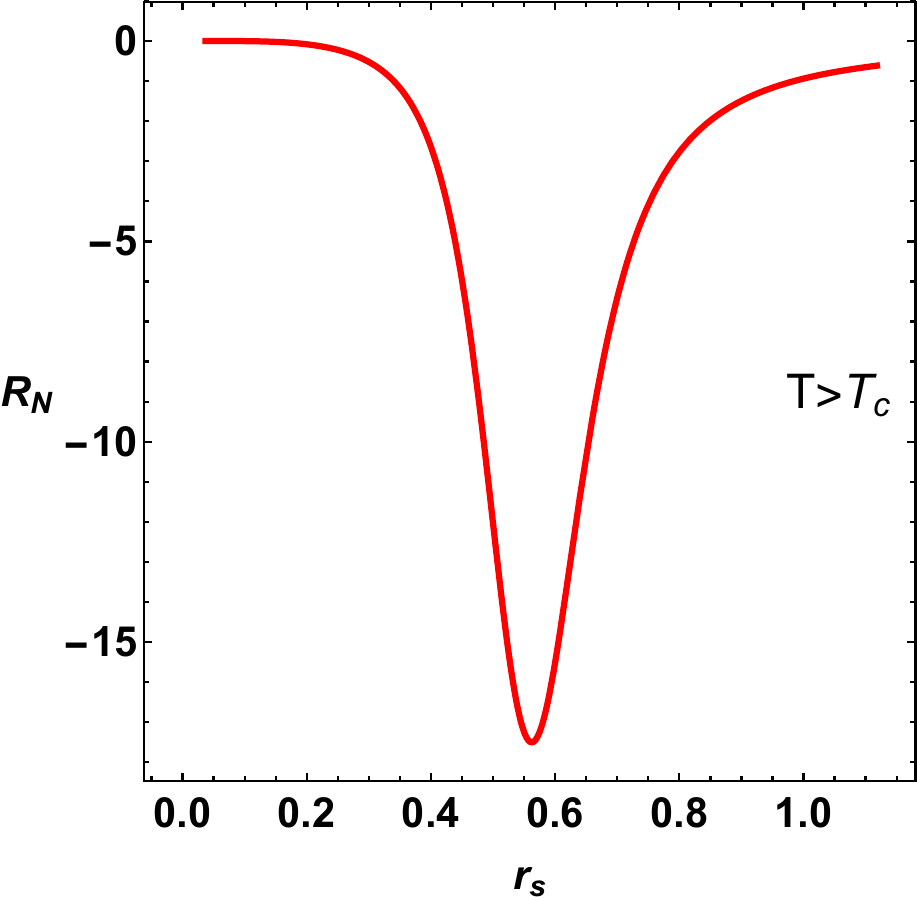}}\par
  \caption{The variation of $R_N$ in terms of the shadow radius $r_s$  with $q=0.1$. The first horizontal side stands for the case of four equal charges, and the second for three equal charges.}
  \label{Fig9}}
 \end{figure*}

  Showing the normalized curvature singularity as a function of the shadow radius is mainly an extension of the study \cite{54,54a,54b} using the Ruppeiner formalism \cite{Ru,Ru1}. In addition, it could be similar to the exploitation of other thermodynamic geometry structures, such as Weinhold \cite{Wei} and Quevedo \cite{Que} formalisms. Since the shadow radius and the horizon radius are correlated, this implies a similarity in the behavior of the normalized curvature singularity either in terms of the function of the horizon radius or in terms of the shadow radius. \textcolor{black}{From Fig.~$\ref{Fig9}$, we can see that, in contrast to the case of the charged AdS black hole~\cite{Wei:2015iwa}, $R_N$ is always negative. This behavior is similar to the Van der Waals fluid system case, implying that interactions among the microstructures of the four $\mathbb{S}\mathbb{T}\mathbb{U}$ black holes are always attractive for the considered charge configurations.  At a closer look, it is observed that three regimes can be distinguished, depending on the value of the temperature. For $T<T_c$, there exist two critical points. $r_{s1}$ and $r_{s2}$, respectively, where the normalized curvature $R_N$ diverges. The first point lies in the small black hole phase (i.e. at small $r_s$), while the second in the large phase (i.e. at large $r_s$). These two points gradually get closer as $T$ increases, and finally collapse as $T$ reaches the critical value $T=T_c$. Lastly, the normalized scalar curvature has no divergent points at higher temperature $T>T_c$. Again, such a behavior is similar to the Van der Waals fluid, in which no divergent points occurs above the critical temperature. Interestingly enough, the same features of the scalar curvature have been found in~\cite{Gogoi:2021syo} for the case of five-dimensional R-charged black hole in extended thermodynamic space.
  }

As a final note, let us remark that a relevant quantity to explore the global stability and phase transitions of black holes is the Gibbs free energy $G$~\cite{20,Sari,Sari2,Sari3}, which is defined as the enthalpy (or mass) minus the product of the temperature times the entropy of the system. Indeed, below the critical pressure, the $G-T$ diagram exhibits a swallow tail behavior, which is typical of first-order phase transitions. However, because of the non-trivial form of the characteristic thermodynamic quantities for $\mathbb{S}\mathbb{T}\mathbb{U}$ black holes (see Eqs.~\eqref{Eq7}-\eqref{Eq12}), computations appear quite cumbersome and will be reserved for future investigation.

\section{Conclusion}
In this work, we have studied the shadow behavior and energy emission rate pertinent to $\mathbb{S}\mathbb{T}\mathbb{U}$ black holes with four electric charges. Furthermore, we  have investigated the phase structure using shadow analysis. The findings have demonstrated  that the electric charge affects the shadow black hole in such a way that the shadow's size increases as the electric charge values increase. Indeed, this effect is observed in all the considered charge configurations. Also, it has been shown that the parameter space $\mathcal{M}_q\left(q_1,q_2,q_3,q_4\right)$ influences the size of the shadow black hole in such a way that once the number of the electric charges is reduced, the size decreases. On the other side, the finding about the energy emission rate in relation to the charge parameter have revealed that the energy emission rate decreases with increases in the charge parameter. Also, the parameter space of the four electric charges has a non-trivial impact the energy emission rate. The energy emission rate process has revealed a scenario in which the black hole evaporation process is slower and has an extended lifetime. The approach to the phase structure of the $\mathbb{S}\mathbb{T}\mathbb{U}$ black hole has been taken into account from the point of view of the shadow. As a consequence, the $T$-$r_s$ phase transition clearly demonstrates similarity with the ordinary phase transition $T$-$r_h$. We have inspected the isobaric $T$-$r_s$ curve with respect to two cases of charge configuration. Furthermore, the heat capacity has been  used to claim the local stability of the black as a function of the shadow radius. The last task has been carried out to reveal the microstructure behavior using the Ruppeiner formalism in terms of shadow analysis. We stress that a similar analysis can be conducted within Weinhold \cite{Wei} and Quevedo \cite{Que} frameworks, which provide an effort to extract the microscopic interaction information of black holes from the axioms
of thermodynamics.
Therefore, our research underscores the significant effect of both the charge parameter and the parameter space on the shadow, energy emission rate, and phase structure properties of the $\mathbb{S}\mathbb{T}\mathbb{U}$ black hole.

This work comes up with certain issues and questions for the purpose of discovering the shadow behaviors of the $\mathbb{S}\mathbb{T}\mathbb{U}$ black holes. And this work will be a topic for revealing the quasinormal mode, deflection angle, and accretion disk. More  theoretical and phenomenological effort is inevitably required along these directions, especially in view of the release of new experimental data that promise to give unparalleled insight into black hole physics beyond Einstein's gravity.

\revappendix
\section{Black hole sparsity}
\label{A2}
We here present detailed computations of black hole sparsity for the four cases of charge configurations considered in Sec.~\ref{S3}. Applying the definition~\eqref{defSpars}, we find:
    \begin{itemize}
    \item case 1: $q_1=q_2=q_3=q_4=q$,
        \begin{equation}
 \eta  = \frac{64 \pi ^3 r_h^2 \left(r_h+q\right){}^2}{27 \left(r_h^2-g^2 \left(q-3 r_h\right)
   \left(r_h+q\right){}^3\right){}^2}.
\end{equation}
\item case 2: $q_1=q_2=q_3=q$ and $q_4=0$,
\begin{align}
   \eta  &=  \frac{\eta_1}{\eta_2},
\end{align}
with
\begin{widetext}
\begin{align}
    \eta_1 &=64 \pi ^3 \Biggl\{\frac{1}{4} \sqrt{q^2+\frac{2\ 2^{2/3} \left(\sqrt{768 \pi ^6 r_h^3 \left(q+r_h\right){}^9+81 \pi ^6 q^4 r_h^2 \left(q+r_h\right){}^6}-9 \pi ^3 q^2 r_h \left(q+r_h\right){}^3\right){}^{2/3}-16 \sqrt[3]{6} \pi ^2 r_h \left(q+r_h\right){}^3}{3^{2/3} \pi  \sqrt[3]{\sqrt{768 \pi ^6 r_h^3 \left(q+r_h\right){}^9+81 \pi ^6 q^4 r_h^2
   \left(q+r_h\right){}^6}-9 \pi ^3 q^2 r_h \left(q+r_h\right){}^3}}}\nonumber\\
   &+\frac{1}{2} \Bigg(\frac{q^3}{2 \sqrt{q^2+\frac{2\ 2^{2/3} \left(\sqrt{768 \pi ^6 r_h^3 \left(q+r_h\right){}^9+81 \pi ^6 q^4 r_h^2 \left(q+r_h\right){}^6}-9 \pi ^3 q^2 r_h \left(q+r_h\right)^3\right){}^{2/3}-16 \sqrt[3]{6} \pi ^2 r_h (q+r_h)^3}{3^{2/3} \pi  \sqrt[3]{\sqrt{768 \pi ^6 r_h^3
   \left(q+r_h\right){}^9+81 \pi ^6 q^4 r_h^2 \left(q+r_h\right){}^6}-9 \pi ^3 q^2 r_h \left(q+r_h\right){}^3}}}}+\frac{q^2}{2}\nonumber\\
   &-\frac{\sqrt[3]{\sqrt{768 \pi ^6 r_h^3 \left(q+r_h\right)^9+81 \pi ^6 q^4 r_h^2 \left(q+r_h\right)^6}-9 \pi ^3 q^2 r_h \left(q+r_h\right)^3}}{\sqrt[3]{2} 3^{2/3} \pi }\nonumber\\
   &+\frac{4 \sqrt[3]{\frac{2}{3}} \pi  r_h \left(q+r_h\right){}^3}{\sqrt[3]{\sqrt{768
   \pi ^6 r_h^3 \left(q+r_h\right){}^9+81 \pi ^6 q^4 r_h^2 \left(q+r_h\right){}^6}-9 \pi ^3 q^2 r_h \left(q+r_h\right){}^3}}\Bigg)^{1/2}\Bigg)^{5/2} \Bigg(\frac{q}{4}\nonumber\\
   &+\frac{1}{2} \Bigg(\frac{q^3}{2 \sqrt{q^2+\frac{2\ 2^{2/3} \left(\sqrt{768 \pi ^6 r_h^3 \left(q+r_h\right)^9+81 \pi ^6 q^4 r_h^2 \left(q+r_h\right)^6}-9 \pi ^3 q^2 r_h \left(q+r_h\right)^3\right){}^{2/3}-16 \sqrt[3]{6}
   \pi ^2 r_h \left(q+r_h\right){}^3}{3^{2/3} \pi  \sqrt[3]{\sqrt{768 \pi ^6 r_h^3 \left(q+r_h\right){}^9+81 \pi ^6 q^4 r_h^2 \left(q+r_h\right){}^6}-9 \pi ^3 q^2 r_h \left(q+r_h\right){}^3}}}}+\frac{q^2}{2}\nonumber\\
   &-\frac{\sqrt[3]{\sqrt{768 \pi ^6 r_h^3 \left(q+r_h\right){}^9+81 \pi ^6 q^4 r_h^2 \left(q+r_h\right){}^6}-9 \pi ^3 q^2 r_h \left(q+r_h\right){}^3}}{\sqrt[3]{2} 3^{2/3} \pi
   }\nonumber\\
   &+\frac{4 \sqrt[3]{\frac{2}{3}} \pi  r_h \left(q+r_h\right){}^3}{\sqrt[3]{\sqrt{768 \pi ^6 r_h^3 \left(q+r_h\right){}^9+81 \pi ^6 q^4 r_h^2 \left(q+r_h\right){}^6}-9 \pi ^3 q^2 r_h \left(q+r_h\right)^3}}\Bigg)^{1/2} -\frac{3 q}{4}\nonumber\\
   &+\frac{1}{4} \sqrt{q^2+\frac{2\ 2^{2/3} \left(\sqrt{768 \pi ^6 r_h^3 \left(q+r_h\right){}^9+81 \pi ^6 q^4 r_h^2 \left(q+r_h\right){}^6}-9 \pi ^3 q^2 r_h
   \left(q+r_h\right){}^3\right){}^{2/3}-16 \sqrt[3]{6} \pi ^2 r_h \left(q+r_h\right){}^3}{3^{2/3} \pi  \sqrt[3]{\sqrt{768 \pi ^6 r_h^3 \left(q+r_h\right){}^9+81 \pi ^6 q^4 r_h^2 \left(q+r_h\right){}^6}-9 \pi ^3 q^2 \left(q+r_h\right){}^3 r_h}}}\Biggr\}^{3/2}\nonumber
   \end{align}
   and
   \begin{align}
   \eta_2& =27 \Biggl\{\Bigg(3 \Bigg(\frac{1}{4}
   \sqrt{q^2+\frac{2\ 2^{2/3} \left(\sqrt{768 \pi ^6 r_h^3 \left(q+r_h\right){}^9+81 \pi ^6 q^4 r_h^2
   \left(q+r_h\right){}^6}-9 \pi ^3 q^2 r_h \left(q+r_h\right){}^3\right){}^{2/3}-16 \sqrt[3]{6} \pi ^2 r_h
   \left(q+r_h\right){}^3}{3^{2/3} \pi  \sqrt[3]{\sqrt{768 \pi ^6 r_h^3 \left(q+r_h\right){}^9+81 \pi ^6 q^4 r_h^2
   \left(q+r_h\right){}^6}-9 \pi ^3 q^2 r_h \left(q+r_h\right){}^3}}}\nonumber\\
   &+\frac{1}{2} \Bigg(\frac{q^3}{2
   \sqrt{q^2+\frac{2\ 2^{2/3} \left(\sqrt{768 \pi ^6 r_h^3 \left(q+r_h\right){}^9+81 \pi ^6 q^4 r_h^2
   \left(q+r_h\right){}^6}-9 \pi ^3 q^2 r_h \left(q+r_h\right){}^3\right){}^{2/3}-16 \sqrt[3]{6} \pi ^2 r_h
   \left(q+r_h\right){}^3}{3^{2/3} \pi  \sqrt[3]{\sqrt{768 \pi ^6 r_h^3 \left(q+r_h\right){}^9+81 \pi ^6 q^4 r_h^2
   \left(q+r_h\right){}^6}-9 \pi ^3 q^2 r_h \left(q+r_h\right){}^3}}}}+\frac{q^2}{2}\nonumber\\
   &-\frac{\sqrt[3]{\sqrt{768 \pi ^6
   r_h^3 \left(q+r_h\right){}^9+81 \pi ^6 q^4 r_h^2 \left(q+r_h\right){}^6}-9 \pi ^3 q^2 r_h
   \left(q+r_h\right){}^3}}{\sqrt[3]{2} 3^{2/3} \pi }\nonumber\\
   &+\frac{4 \sqrt[3]{\frac{2}{3}} \pi  r_h
   \left(q+r_h\right){}^3}{\sqrt[3]{\sqrt{768 \pi ^6 r_h^3 \left(q+r_h\right){}^9+81 \pi ^6 q^4 r_h^2
   \left(q+r_h\right){}^6}-9 \pi ^3 q^2 r_h \left(q+r_h\right){}^3}}\Bigg)^{1/2} \Bigg){}^4+6 q \Bigg(-\frac{3 q}{4}\nonumber\\
   &+\frac{1}{4}
   \sqrt{q^2+\frac{2\ 2^{2/3} \left(\sqrt{768 \pi ^6 r_h^3 \left(q+r_h\right){}^9+81 \pi ^6 q^4 r_h^2
   \left(q+r_h\right){}^6}-9 \pi ^3 q^2 r_h \left(q+r_h\right){}^3\right){}^{2/3}-16 \sqrt[3]{6} \pi ^2 r_h
   \left(q+r_h\right){}^3}{3^{2/3} \pi  \sqrt[3]{\sqrt{768 \pi ^6 r_h^3 \left(q+r_h\right){}^9+81 \pi ^6 q^4 r_h^2
   \left(q+r_h\right){}^6}-9 \pi ^3 q^2 r_h \left(q+r_h\right){}^3}}}\nonumber\\
   &+\frac{1}{2} \Bigg(\frac{q^3}{2
   \sqrt{q^2+\frac{2\ 2^{2/3} \left(\sqrt{768 \pi ^6 r_h^3 \left(q+r_h\right){}^9+81 \pi ^6 q^4 r_h^2
   \left(q+r_h\right){}^6}-9 \pi ^3 q^2 r_h \left(q+r_h\right){}^3\right){}^{2/3}-16 \sqrt[3]{6} \pi ^2 r_h
   \left(q+r_h\right){}^3}{3^{2/3} \pi  \sqrt[3]{\sqrt{768 \pi ^6 r_h^3 \left(q+r_h\right){}^9+81 \pi ^6 q^4 r_h^2
   \left(q+r_h\right){}^6}-9 \pi ^3 q^2 r_h \left(q+r_h\right){}^3}}}}+\frac{q^2}{2}\nonumber
   \\
   &-\frac{\sqrt[3]{\sqrt{768 \pi ^6
   r_h^3 \left(q+r_h\right){}^9+81 \pi ^6 q^4 r_h^2 \left(q+r_h\right){}^6}-9 \pi ^3 q^2 r_h
   \left(q+r_h\right){}^3}}{\sqrt[3]{2} 3^{2/3} \pi }\nonumber\\
   &+\frac{4 \sqrt[3]{\frac{2}{3}} \pi  r_h
   \left(q+r_h\right){}^3}{\sqrt[3]{\sqrt{768 \pi ^6 r_h^3 \left(q+r_h\right){}^9+81 \pi ^6 q^4 r_h^2
   \left(q+r_h\right){}^6}-9 \pi ^3 q^2 r_h \left(q+r_h\right){}^3}}\Bigg)^{1/2} \Bigg){}^3+3 q^2 \Bigg(-\frac{3
   q}{4}\nonumber\\
   &+\frac{1}{4} \sqrt{q^2+\frac{2\ 2^{2/3} \left(\sqrt{768 \pi ^6 r_h^3 \left(q+r_h\right){}^9+81 \pi ^6 q^4
   r_h^2 \left(q+r_h\right){}^6}-9 \pi ^3 q^2 r_h \left(q+r_h\right){}^3\right){}^{2/3}-16 \sqrt[3]{6} \pi ^2 r_h
   \left(q+r_h\right){}^3}{3^{2/3} \pi  \sqrt[3]{\sqrt{768 \pi ^6 r_h^3 \left(q+r_h\right){}^9+81 \pi ^6 q^4 r_h^2
   \left(q+r_h\right){}^6}-9 \pi ^3 q^2 r_h \left(q+r_h\right){}^3}}}\nonumber\\
   &+\frac{1}{2} \Bigg(\frac{q^3}{2
   \sqrt{q^2+\frac{2\ 2^{2/3} \left(\sqrt{768 \pi ^6 r_h^3 \left(q+r_h\right){}^9+81 \pi ^6 q^4 r_h^2
   \left(q+r_h\right){}^6}-9 \pi ^3 q^2 r_h \left(q+r_h\right){}^3\right){}^{2/3}-16 \sqrt[3]{6} \pi ^2 r_h
   \left(q+r_h\right){}^3}{3^{2/3} \pi  \sqrt[3]{\sqrt{768 \pi ^6 r_h^3 \left(q+r_h\right){}^9+81 \pi ^6 q^4 r_h^2
   \left(q+r_h\right){}^6}-9 \pi ^3 q^2 r_h \left(q+r_h\right){}^3}}}}+\frac{q^2}{2}\nonumber\\
   &-\frac{\sqrt[3]{\sqrt{768 \pi ^6
   r_h^3 \left(q+r_h\right){}^9+81 \pi ^6 q^4 r_h^2 \left(q+r_h\right){}^6}-9 \pi ^3 q^2 r_h
   \left(q+r_h\right){}^3}}{\sqrt[3]{2} 3^{2/3} \pi }\nonumber\\
   &+\frac{4 \sqrt[3]{\frac{2}{3}} \pi  r_h
   \left(q+r_h\right){}^3}{\sqrt[3]{\sqrt{768 \pi ^6 r_h^3 \left(q+r_h\right){}^9+81 \pi ^6 q^4 r_h^2
   \left(q+r_h\right){}^6}-9 \pi ^3 q^2 r_h \left(q+r_h\right){}^3}}\Bigg)^{1/2} \Bigg){}^2\Bigg) g^2+\Bigg(-\frac{3
   q}{4}\nonumber\\
   &+\frac{1}{4} \sqrt{q^2+\frac{2\ 2^{2/3} \left(\sqrt{768 \pi ^6 r_h^3 \left(q+r_h\right){}^9+81 \pi ^6 q^4
   r_h^2 \left(q+r_h\right){}^6}-9 \pi ^3 q^2 r_h \left(q+r_h\right){}^3\right){}^{2/3}-16 \sqrt[3]{6} \pi ^2 r_h
   \left(q+r_h\right){}^3}{3^{2/3} \pi  \sqrt[3]{\sqrt{768 \pi ^6 r_h^3 \left(q+r_h\right){}^9+81 \pi ^6 q^4 r_h^2
   \left(q+r_h\right){}^6}-9 \pi ^3 q^2 r_h \left(q+r_h\right){}^3}}}\nonumber\\
   &+\frac{1}{2} \Bigg(\frac{q^3}{2
   \sqrt{q^2+\frac{2\ 2^{2/3} \left(\sqrt{768 \pi ^6 r_h^3 \left(q+r_h\right){}^9+81 \pi ^6 q^4 r_h^2
   \left(q+r_h\right){}^6}-9 \pi ^3 q^2 r_h \left(q+r_h\right){}^3\right)^{2/3}-16 \sqrt[3]{6} \pi ^2 r_h
   \left(q+r_h\right){}^3}{3^{2/3} \pi  \sqrt[3]{\sqrt{768 \pi ^6 r_h^3 \left(q+r_h\right){}^9+81 \pi ^6 q^4 r_h^2
   \left(q+r_h\right){}^6}-9 \pi ^3 q^2 r_h \left(q+r_h\right){}^3}}}}+\frac{q^2}{2}\nonumber\\
   &-\frac{\sqrt[3]{\sqrt{768 \pi ^6
   r_h^3 \left(q+r_h\right){}^9+81 \pi ^6 q^4 r_h^2 \left(q+r_h\right){}^6}-9 \pi ^3 q^2 r_h
   \left(q+r_h\right){}^3}}{\sqrt[3]{2} 3^{2/3} \pi }\nonumber\\
   &+\frac{4 \sqrt[3]{\frac{2}{3}} \pi  r_h
   \left(q+r_h\right){}^3}{\sqrt[3]{\sqrt{768 \pi ^6 r_h^3 \left(q+r_h\right){}^9+81 \pi ^6 q^4 r_h^2
   \left(q+r_h\right){}^6}-9 \pi ^3 q^2 r_h \left(q+r_h\right){}^3}}\Bigg)^{1/2} \Bigg)^2\Biggr\}^2.\nonumber
\end{align}
\end{widetext}
\item case 3: $q_1=q_2=q$ and $q_3=q_4=0$, 
\begin{widetext}
    \begin{equation}
       \eta= \frac{128 \pi ^4 \left(\sqrt{\pi } q \sqrt{4 \pi  r_h \left(r_h+q\right)+\pi  q^2}+2 \pi  r_h \left(r_h+q\right)+\pi  q^2\right)}{27 r_h \left(r_h+q\right) \left(\sqrt{\pi } g^2 q \sqrt{4 \pi  r_h
   \left(r_h+q\right)+\pi  q^2}+6 \pi  g^2 r_h \left(r_h+q\right)+\pi  \left(g^2 q^2+2\right)\right)^2}.
    \end{equation}
\end{widetext}
\item case 4: $q_1=q$ and $q_2=q_3=q_4=0$ ,
    \begin{equation}
        \eta= \frac{\eta_3}{\eta_4},
    \end{equation}
with
\begin{widetext}
    \begin{align}
        \eta_3&= 64 \pi ^3 \Bigg(-\frac{q}{4}-\frac{1}{4} \Bigg(-\frac{16 \sqrt[3]{\frac{2}{3}} \pi  \left(q+r_h\right) r_h^3}{\sqrt[3]{\sqrt{768 \pi ^6 \left(q+r_h\right){}^3 r_h^9+81 \pi ^6 q^4 \left(q+r_h\right){}^2
   r_h^6}-9 \pi ^3 q^2 r_h^3 \left(q+r_h\right)}}+q^2\nonumber\\
   &+\frac{2 \left(\frac{2}{3}\right)^{2/3} \sqrt[3]{\sqrt{768 \pi ^6 \left(q+r_h\right){}^3 r_h^9+81 \pi ^6 q^4 \left(q+r_h\right){}^2 r_h^6}-9 \pi ^3
   q^2 r_h^3 \left(q+r_h\right)}}{\pi }\Bigg)^2\nonumber\\
   &+\frac{1}{2} \Bigg(\frac{q^3}{2 \sqrt{-\frac{16 \sqrt[3]{\frac{2}{3}} \pi  \left(q+r_h\right) r_h^3}{\sqrt[3]{\sqrt{768 \pi ^6 \left(q+r_h\right){}^3 r_h^9+81 \pi
   ^6 q^4 \left(q+r_h\right){}^2 r_h^6}-9 \pi ^3 q^2 r_h^3 \left(q+r_h\right)}}+q^2+\frac{2 \left(\frac{2}{3}\right)^{2/3} \sqrt[3]{\sqrt{768 \pi ^6 \left(q+r_h\right){}^3 r_h^9+81 \pi ^6 q^4
   \left(q+r_h\right){}^2 r_h^6}-9 \pi ^3 q^2 r_h^3 \left(q+r_h\right)}}{\pi }}}\nonumber\\
   &+\frac{q^2}{2}-\frac{\sqrt[3]{\sqrt{768 \pi ^6 \left(q+r_h\right){}^3 r_h^9+81 \pi ^6 q^4 \left(q+r_h\right){}^2 r_h^6}-9
   \pi ^3 q^2 r_h^3 \left(q+r_h\right)}}{\sqrt[3]{2} 3^{2/3} \pi }\nonumber\\
   &+\frac{4 \sqrt[3]{\frac{2}{3}} \pi  r_h^3 \left(q+r_h\right)}{\sqrt[3]{\sqrt{768 \pi ^6 \left(q+r_h\right){}^3 r_h^9+81 \pi ^6 q^4
   \left(q+r_h\right){}^2 r_h^6}-9 \pi ^3 q^2 r_h^3 \left(q+r_h\right)}}\Bigg)^{1/2}\Bigg)^{7/2} \Bigg(\frac{3 q}{4}\nonumber\\
   &-\frac{1}{4} \Bigg(-\frac{16 \sqrt[3]{\frac{2}{3}} \pi  \left(q+r_h\right)
   r_h^3}{\sqrt[3]{\sqrt{768 \pi ^6 \left(q+r_h\right){}^3 r_h^9+81 \pi ^6 q^4 \left(q+r_h\right){}^2 r_h^6}-9 \pi ^3 q^2 r_h^3 \left(q+r_h\right)}}+q^2\nonumber\\
   &+\frac{2 \left(\frac{2}{3}\right)^{2/3}
   \sqrt[3]{\sqrt{768 \pi ^6 \left(q+r_h\right){}^3 r_h^9+81 \pi ^6 q^4 \left(q+r_h\right){}^2 r_h^6}-9 \pi ^3 q^2 r_h^3 \left(q+r_h\right)}}{\pi }\Bigg)^ 
{1/2} +\frac{1}{2} \Bigg(\frac{q^3}{u}\Bigg)^{1/2}+\frac{q^2}{2}\nonumber\\
&-\frac{\sqrt[3]{\sqrt{768 \pi ^6 \left(q+r_h\right){}^3 r_h^9+81 \pi ^6 q^4 \left(q+r_h\right){}^2 r_h^6}-9 \pi ^3 q^2 r_h^3 \left(q+r_h\right)}}{\sqrt[3]{2}
   3^{2/3} \pi }\nonumber\\
   &+\frac{4 \sqrt[3]{\frac{2}{3}} \pi  r_h^3 \left(q+r_h\right)}{\sqrt[3]{\sqrt{768 \pi ^6 \left(q+r_h\right){}^3 r_h^9+81 \pi ^6 q^4 \left(q+r_h\right){}^2 r_h^6}-9 \pi ^3 q^2 r_h^3\left(q+r_h\right)}}\Bigg)^{1/2}\Bigg)^{1/2}\nonumber
   \end{align}
   and
   \begin{align}
   \eta_4&= 27 \Biggl\{\Bigg(3 \Bigg(\frac{q}{4}+\frac{1}{4} \Bigg(-\frac{16 \sqrt[3]{\frac{2}{3}} \pi  \left(q+r_h\right) r_h^3}{\sqrt[3]{\sqrt{768 \pi ^6 \left(q+r_h\right){}^3 r_h^9+81 \pi ^6 q^4
   \left(q+r_h\right){}^2 r_h^6}-9 \pi ^3 q^2 r_h^3 \left(q+r_h\right)}}+q^2\nonumber\\
   &+\frac{2 \left(\frac{2}{3}\right)^{2/3} \sqrt[3]{\sqrt{768 \pi ^6 \left(q+r_h\right){}^3 r_h^9+81 \pi ^6 q^4
   \left(q+r_h\right){}^2 r_h^6}-9 \pi ^3 q^2 r_h^3 \left(q+r_h\right)}}{\pi }\Bigg)^{1/2}-\frac{1}{2} \Bigg(\frac{q^3}{v}+\frac{q^2}{2}\nonumber\\
   &-\frac{\sqrt[3]{\sqrt{768 \pi ^6 \left(q+r_h\right){}^3 r_h^9+81
   \pi ^6 q^4 \left(q+r_h\right){}^2 r_h^6}-9 \pi ^3 q^2 r_h^3 \left(q+r_h\right)}}{\sqrt[3]{2} 3^{2/3} \pi }\nonumber\\
   &+\frac{4 \sqrt[3]{\frac{2}{3}} \pi  r_h^3 \left(q+r_h\right)}{\sqrt[3]{\sqrt{768 \pi ^6
   \left(q+r_h\right){}^3 r_h^9+81 \pi ^6 q^4 \left(q+r_h\right){}^2 r_h^6}-9 \pi ^3 q^2 r_h^3 \left(q+r_h\right)}}\Bigg)^{1/2}\Bigg){}^4+2 q \Bigg(-\frac{q}{4}\nonumber\\
   &-\frac{1}{4} \Bigg(-\frac{16 \sqrt[3]{\frac{2}{3}}
   \pi  \left(q+r_h\right) r_h^3}{\sqrt[3]{\sqrt{768 \pi ^6 \left(q+r_h\right){}^3 r_h^9+81 \pi ^6 q^4 \left(q+r_h\right){}^2 r_h^6}-9 \pi ^3 q^2 r_h^3 \left(q+r_h\right)}}+q^2\nonumber\\
   &+\frac{2
   \left(\frac{2}{3}\right)^{2/3} \sqrt[3]{\sqrt{768 \pi ^6 \left(q+r_h\right){}^3 r_h^9+81 \pi ^6 q^4 \left(q+r_h\right){}^2 r_h^6}-9 \pi ^3 q^2 r_h^3 \left(q+r_h\right)}}{\pi }\Bigg)^{1/2} +\frac{1}{2}
   \Bigg(\frac{q^3}{v}+\frac{q^2}{2}\nonumber\\
   &-\frac{\sqrt[3]{\sqrt{768 \pi ^6 \left(q+r_h\right){}^3 r_h^9+81 \pi ^6 q^4 \left(q+r_h\right){}^2 r_h^6}-9 \pi ^3 q^2 r_h^3 \left(q+r_h\right)}}{\sqrt[3]{2}
   3^{2/3} \pi }\nonumber\\
   &+\frac{4 \sqrt[3]{\frac{2}{3}} \pi  r_h^3 \left(q+r_h\right)}{\sqrt[3]{\sqrt{768 \pi ^6 \left(q+r_h\right){}^3 r_h^9+81 \pi ^6 q^4 \left(q+r_h\right){}^2 r_h^6}-9 \pi ^3 q^2 r_h^3
   \left(q+r_h\right)}}\Bigg)^{1/2} \Bigg){}^3\Bigg) g^2+\Bigg(\frac{q}{4}+\frac{1}{4} \Bigg(q^2\nonumber\\
   &+\frac{2 \left(\frac{2}{3}\right)^{2/3} \sqrt[3]{\sqrt{768 \pi ^6 \left(q+r_h\right){}^3 r_h^9+81 \pi ^6
   q^4 \left(q+r_h\right){}^2 r_h^6}-9 \pi ^3 q^2 r_h^3 \left(q+r_h\right)}}{\pi }\Bigg)^{1/2}\nonumber\\
   &-\frac{1}{2} \Bigg(\frac{q^3}{2 \sqrt{-\frac{16 \sqrt[3]{\frac{2}{3}} \pi  \left(q+r_h\right) r_h^3}{\sqrt[3]{\sqrt{768
   \pi ^6 \left(q+r_h\right){}^3 r_h^9+81 \pi ^6 q^4 \left(q+r_h\right){}^2 r_h^6}-9 \pi ^3 q^2 r_h^3 \left(q+r_h\right)}}+q^2+\frac{2 \left(\frac{2}{3}\right)^{2/3} \sqrt[3]{\sqrt{768 \pi ^6
   \left(q+r_h\right){}^3 r_h^9+81 \pi ^6 q^4 \left(q+r_h\right){}^2 r_h^6}-9 \pi ^3 q^2 r_h^3 \left(q+r_h\right)}}{\pi }}}\nonumber\\
   &+\frac{q^2}{2}-\frac{\sqrt[3]{\sqrt{768 \pi ^6 \left(q+r_h\right){}^3 r_h^9+81
   \pi ^6 q^4 \left(q+r_h\right){}^2 r_h^6}-9 \pi ^3 q^2 r_h^3 \left(q+r_h\right)}}{\sqrt[3]{2} 3^{2/3} \pi }\nonumber\\
   &+\frac{4 \sqrt[3]{\frac{2}{3}} \pi  r_h^3 \left(q+r_h\right)}{\sqrt[3]{\sqrt{768 \pi ^6
   \left(q+r_h\right){}^3 r_h^9+81 \pi ^6 q^4 \left(q+r_h\right){}^2 r_h^6}-9 \pi ^3 q^2 r_h^3 \left(q+r_h\right)}}\Bigg)^{1/2} \Bigg)^2\Biggr\}^2,\nonumber
    \end{align}
\end{widetext}
where
\begin{widetext}
    \begin{align}
    u&= 2 \Bigg(-\frac{16
   \sqrt[3]{\frac{2}{3}} \pi  \left(q+r_h\right) r_h^3}{\sqrt[3]{\sqrt{768 \pi ^6 \left(q+r_h\right){}^3 r_h^9+81 \pi ^6 q^4 \left(q+r_h\right){}^2 r_h^6}-9 \pi ^3 q^2 r_h^3
   \left(q+r_h\right)}}+q^2\nonumber\\
   &+\frac{2 \left(\frac{2}{3}\right)^{2/3} \sqrt[3]{\sqrt{768 \pi ^6 \left(q+r_h\right){}^3 r_h^9+81 \pi ^6 q^4 \left(q+r_h\right){}^2 r_h^6}-9 \pi ^3 q^2 r_h^3
   \left(q+r_h\right)}}{\pi }\Bigg)\nonumber,\\
   v&=2 \Bigg(-\frac{16 \sqrt[3]{\frac{2}{3}} \pi  \left(q+r_h\right) r_h^3}{\sqrt[3]{\sqrt{768 \pi
   ^6 \left(q+r_h\right){}^3 r_h^9+81 \pi ^6 q^4 \left(q+r_h\right){}^2 r_h^6}-9 \pi ^3 q^2 r_h^3 \left(q+r_h\right)}}+q^2\nonumber\\
   &+\frac{2 \left(\frac{2}{3}\right)^{2/3} \sqrt[3]{\sqrt{768 \pi ^6
   \left(q+r_h\right){}^3 r_h^9+81 \pi ^6 q^4 \left(q+r_h\right){}^2 r_h^6}-9 \pi ^3 q^2 r_h^3 \left(q+r_h\right)}}{\pi }\Bigg)^{1/2}. \nonumber\\
\end{align}
\end{widetext}
\end{itemize}

\section*{Acknowledgments}
The authors are grateful to the anonymous Referees, who contributed to improve the quality of the manuscript with their comments and recommendations. This work was supported by the Ministry of Science and Higher Education of the Republic of Kazakhstan, Grant AP14870191. GGL acknowledges the Spanish ``Ministerio de Universidades'' for the awarded Maria Zambrano fellowship and funding received
from the European Union - NextGenerationEU. 

\section*{Data availability}
No data were generated for the research described in this article.

\bibliographystyle{apsrev4-2}

\end{document}